\documentclass[
article,
aps,
preprint,    
showpacs,
preprintnumbers, 
amsmath,amssymb,
]{revtex4-2}
\usepackage{graphicx}
\usepackage{dcolumn}
\usepackage{bm}
\usepackage[version=3]{mhchem}
\usepackage{color}
\usepackage{rotating}
\usepackage{lineno}

\begin{document}
\title{First-principles insight in structure-property relationships of hexagonal Si and Ge polytypes}%
\author{Martin Keller$^1$, Abderrezak Belabbes$^{1,2}$, J\"urgen Furthm\"uller$^1$, Friedhelm Bechstedt$^1$, and Silvana Botti$^1$}
\affiliation{%
$^1$ Friedrich-Schiller-Universit\"at Jena, Institut f\"ur Festk\"orpertheorie und -optik, Max-Wien-Platz 1, 07743 Jena, Germany\\
$^{2}$ Department of Physics, Sultan Qaboos University, P.O. Box 36, PC 123, Muscat, Oman
}%
\begin{abstract}
Hexagonal SiGe is a promising material for combining electronic and photonic technologies. In this work, the energetic, structural, elastic and electronic properties of the hexagonal polytypes (2$H$, 4$H$ and 6$H$) of silicon and germanium are thoroughly analyzed under equilibrium conditions. For this purpose, we apply state-of-the-art density functional theory. The phase diagram, obtained in the framework of a generalized Ising model, shows that the diamond structure is the most stable under ambient conditions, but hexagonal modifications are close to the phase boundary, especially for Si.
Our band-structure calculations using the MBJLDA and HSE06 exchange correlation functionals predict significant changes in electronic states with hexagonality. While Si crystals are always semiconductors with indirect band gaps, the hexagonal Ge polytypes have direct band gaps. The branch point energies for Ge crystals are below the valence band maxima, and therefore the formation of hole gases on Ge surfaces is favoured. Band alignment based on the branch point energy leads to type-I heterocrystalline interfaces between Ge polytypes, where electrons and holes can be trapped in the layer with the higher hexagonality. In contrast, the energy shift of the indirect conduction band minima of Si polytypes is rather weak, leading to delocalization of excited electrons at interfaces, while only holes can localize in the layer with higher hexagonality.
\end{abstract}
\maketitle

\section{\label{sec1}Introduction}

The group IV elements silicon (Si) and germanium (Ge) crystallize under ambient conditions in a cubic diamond structure. The bonding is characterized by $sp^3$ hybridization and, consequently, the nearest neighbor atoms form regular tetrahedra. However, under higher pressure
several Si and Ge polymorphs with different coordination have been reported \cite{Olijnyk.Holzapfel:1984:JPC,Hu.Merkle.ea:1986:PRB,Duclos.Vohra.ea:1987:PRL,Duclos.Vohra.ea:1990:PRB,McMahon.Nelmes:1993:PRB,Crain.Ackland.ea:1994:PRB,Hanfland.Schwarz.ea:1999:PRL,Takemura.Schwarz.ea:2001:pssb}.
In addition, using various deposition and growth techniques, Si and Ge polytypes that differ from the diamond structure have also been
observed.

\begin{figure}[h]
\includegraphics[width=16cm]{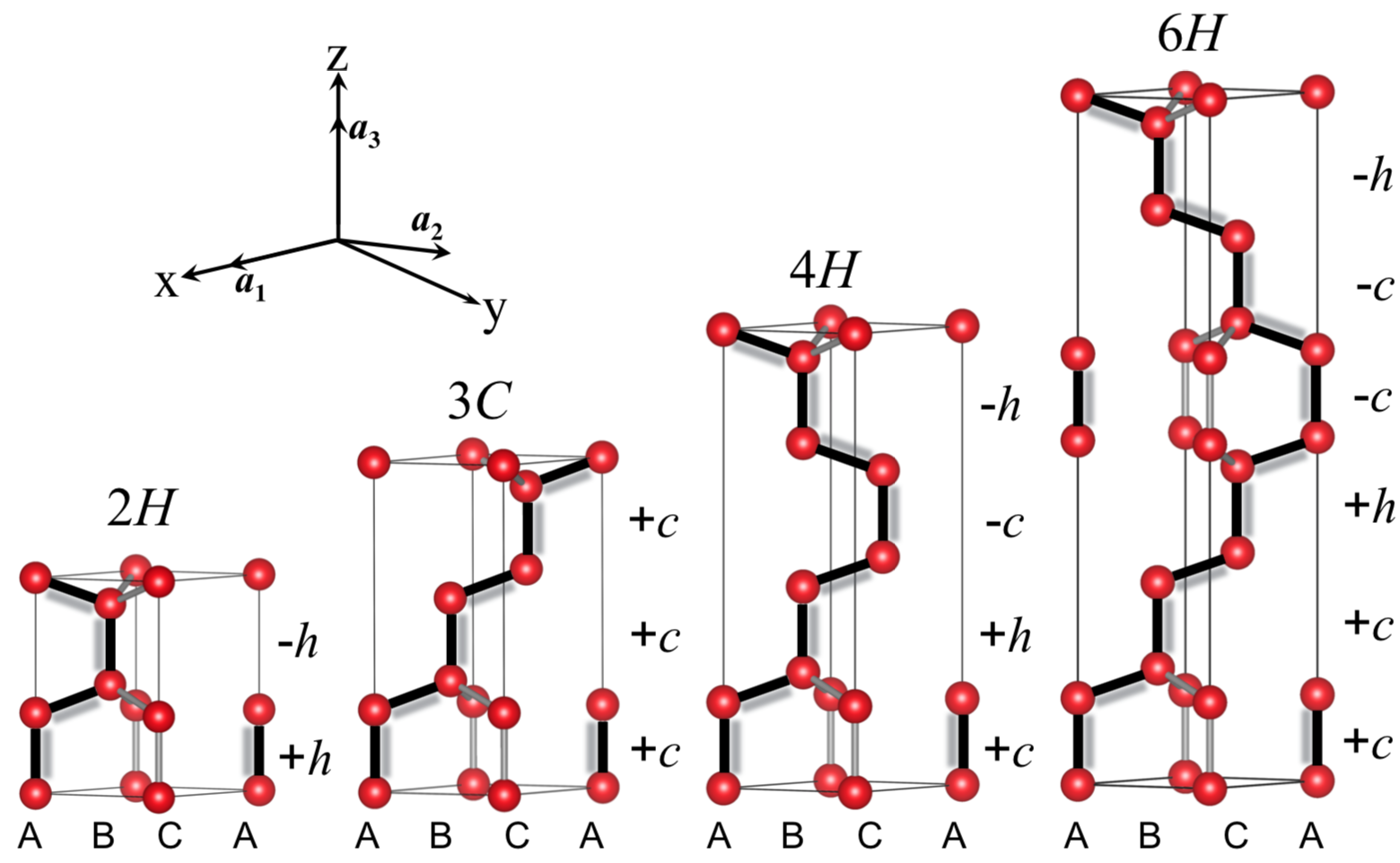} 
\caption{Bond stacking along the [0001] direction in the hexagonal unit cells of the four polytypes. Group-IV atoms
are displayed as red circles. Bonds in a (11$\bar{2}$0) plane are indicated by black solid lines. The cubic ($c$) or
hexagonal ($h$) character of each bilayer is defined by the nonparallel bond in the plane. The signs $+$ and $-$ denote
the orientation of the bilayer. For a more detailed explanation, see \cite{Bechstedt.Kackell.ea:1997:pssb}. The primitive
basis vectors $\bm{a}_i$ ($i=1,2,3$) are also shown.}
\label{fig1}
\end{figure}

Polytypism is a one-dimensional type of polymorphism which occurs, e.g., when the geometry of structural layers is preserved but the number of layers in the layer-stacking sequence is altered. In this setting, hexagonal polytypes of the diamond structure can be formed by
keeping the tetrahedral coordination while varying the stacking sequence along the cubic [111] (corresponding to the hexagonal [0001]) direction
\cite{Verma.Krishna:1966:Book}. As it can be observed in Fig.~\ref{fig1}, the resulting polytypes only differ in the manner how Si or Ge bilayers are oriented along the stacking axis, yielding either a {\it chair} or a {\it boat} conformation. One may equivalently speak about the stacking of {\it eclipsed} or {\it staggered} bilayers \cite{Weltner:1969:JCP}. In Ramsdell's notation \cite{Ramsdell:1947:AM} the stacking in the diamond structure
is purely chair-like (eclipsed) and denoted with 3$C$, since periodicity in [111] direction is reached after three cubic ($C$) bilayers.
The space group is $O^7_h$ ($Fd3m$). The pure stacking of boat-like (staggered) conformers leads to the
hexagonal lonsdaleite 2$H$ structure with space group $D^4_{6h}$ ($P6_3/mmc$) and two hexagonal ($H$) bilayers to reach periodicity
in [0001] direction. In addition, in Fig.~\ref{fig1} two other hexagonal polytypes 4$H$ and 6$H$ with four or six bilayers and, hence, eight
or twelve atoms in their hexagonal unit cell are displayed. The space group of $pH$ polytypes $(p=2,4,6)$ is still $D^4_{6h}$ ($P6_3/mmc$).
The ratio of the number of staggered bilayers to the total number of
such layers gives the percentage of hexagonality $h$ with 0\% for 3$C$, 33\% for 6$H$, 50\% for 4$H$, and
100\% for 2$H$. More complex arrangements of eclipsed and staggered bilayers can lead also to rhombohedral ($R$) polytypes $pR$
with space group $R3m$ ($C^5_{3v}$).

Londsdaleite 2$H$ is the most studied polytype of Si after diamond $3C$ and been synthesized using a variety of methods
\cite{Minomura.Drickamer:1962:JoPaCoS,Zhang.Iqbal.ea:1999:APL,Domnich.Gogotsi.ea:2000:APL,Viera.Mikikian.ea:2002:JoAP,FontcubertaiMorral.Arbiol.ea:2007:AM,Alet.Yu.ea:2008:JMC,Ahn.Kim.ea:2021:SSaT}.
Lonsdaleite silicon is present in core-shell nanowires
\cite{FontcubertaiMorral.Arbiol.ea:2007:AM,Lopez.Hemesath.ea:2009:Nl,Hauge.Verheijen.ea:2015:NL,Dixit.Shukla:2018:JoAP} and 4$H$-Si nanoplatelets have been recently reported \cite{Pandolfi2021}.
The transformation of cubic Si into hexagonal polytpes is also observed for high-temperature indentation~\cite{Eremenko.Nikitenko:1972:pssa}, plastic deformation~\cite{Pirouz.Chaim.ea:1990:AMeM,Dahmen.Westmacott.ea:1990:AMeM}, ion
implantation~\cite{Tan.Foell.ea:1981:PMA}, low-pressure chemical vapor deposition~\cite{Cerva:1991:JoMR}, pulsed laser beam annealing~\cite{Kim.Lee:1996:ML} and nanoribbon growth~\cite{Qiu.Bender.ea:2015:SR5}. It is worthwhile mentioning that the rhombohedral polytype 9$R$ has been observed in Si nanowires \cite{Lopez.Hemesath.ea:2009:Nl,Nicotra.Bongiorno.ea:2011:MP}.
Ribbons of lonsdaleite Ge have been produced in a diamond-Ge matrix already thirty years ago
\cite{Xiao.Pirouz:1992:JoMR,Pirouz.Garg.ea:1995:MOPL,Muellner.Pirouz:1997:MSaEA}. More recently, temperature indentation~\cite{Dushaq.Nayfeh.ea:2019:SR} and ultraviolet laser ablation at low pressure\cite{Zhang.Iqbal.ea:2000:SSC,Haberl.Guthrie.ea:2014:PRB} have been employed to obtain hexagonal phases of Ge. Similarly  to  the case of Si, different routes toward hexagonal Ge nanowires have been explored~\cite{Hauge.Verheijen.ea:2015:NL,Jeon.Dayeh.ea:2013:NL,Vincent.Patriarche.ea:2014:NL,Fadaly:2020:N,Scalise.Sarikov.ea:2021:ASS}. Other hexagonal polytypes beside $2H$, e.g. 4$H$ Ge, have also been reported
\cite{LopezCruz.Cardona:1983:SSC,Parsons.Hoelke:1983:N,Kiefer.Hlukhyy.ea:2010:JMC,Serghiou.Odling.ea:2021:CEJ}.

Although hexagonal polytypes like 2$H$ and 4$H$ of Si and Ge are well documented experimentally, little is known from
theoretical studies about their electronic properties in comparison with the abundant literature for SiC (see, e.g., \cite{Bechstedt.Kackell.ea:1997:pssb}).
One publication~\cite{Raffy.Furthmueller.ea:2002:PRB} addresses their structural, cohesive and electronic properties in the framework of
the density functional theory (DFT) \cite{Hohenberg.Kohn:1964,Kohn.Sham:1965:PR} with the local density approximation~\cite{Kohn.Sham:1965:PR} for the exchange-correlation (XC) functional. The 2$H$ phases of Si and Ge have been recently investigated in detail, including elastic
properties, with DFT calculations using advanced XC functionals for improved band structures~\cite{Roedl.Sander.ea:2015:PRB,Roedl.Furthmueller.ea:2019:PRM,Suckert.Roedl.ea:2021:PRM}. In fact, the Kohn-Sham band structures
obtained from standard functionals in the generalized gradient approximation (GGA) drastically underestimate the fundamental gaps and interband energies of semiconductors
\cite{Bechstedt:2015:Book}, and germanium is wrongly predicted to be metallic independently of the polytype \cite{Cartoixa.Palummo.ea:2017:NL}.
Accurate quasiparticle (QP) band structures can be obtained, for instance, within the GW approximation to the XC self-energy
\cite{Bechstedt:2015:Book}. For 2$H$-Si such GW calculations are available \cite{Roedl.Sander.ea:2015:PRB}.
It has however been shown that more efficient computational approaches allow to obtain QP bandstructures of the same quality at a lower computational cost~\cite{Roedl.Furthmueller.ea:2019:PRM,Wang.Zhang.ea:2021:APL,Bao.Guo.ea:2021:JAP,Kaewmaraya.Vincent.ea:2017:JPCC}.
The most reliable of such approaches consist in applying hybrid XC functionals, e.g. the Heyd-Scuseria-Ernzerhof (HSE06) functional~\cite{Heyd.Scuseria.ea:2003,Heyd.Scuseria.ea:2006:JoCP} or
a meta-GGA XC functional called MBJLDA \cite{Bao.Guo.ea:2021:JAP,Laubscher.Kuefner.ea:2015:JoPCM,Waroquiers.Lherbier.ea:2013:PRB}.
The MBJLDA functional is based on a modified Becke-Johnson (MBJ) exchange potential
\cite{Becke.Johnson:2006:TJoCP,Tran.Blaha.ea:2007:JoPCM,Tran.Blaha:2009:PRL} together with LDA correlation. We note that an
empirical-pseudopotential method (EMP) has been also applied in literature to the 2$H$ polytypes
\cite{Joannopoulos.Cohen:1973:PRB,Joannopoulos.Cohen:1973:PRBa,De.Pryor:2014:JoPCM}.

In this work, we present results for ground-state and excited-state properties of 2$H$, 4$H$ and 6$H$ hexagonal polytypes of the
group-IV elements Si and Ge. We include also calculations for the 3$C$ diamond phase, using  a non-primitive
hexagonal unit cell to facilitate the comparison. 
For the ground state we discuss atomic structures, phase stability and the elastic coefficients. Among the excited-state properties we focus on the modification of band structures and band gaps with hexagonality. We further calculate branch-point
energies and apply them to obtain band alignments at the interface between different polytypes.

\section{\label{sec2}Theoretical and computational methods}
\subsection{\label{sec2a}Structural properties}

All calculations were performed with the Vienna Ab-initio Simulation Package (VASP)
\cite{Kresse.Furthmueller:1996:PRB,Kresse.Furthmuller:1996:CMS}. The wavefunctions and pseudopotentials are described within
the projector-augmented wave method \cite{Kresse.Joubert:1999:PRB} and the shallow Ge3$d$ electrons are considered as valence electrons. The plane-wave expansion is restricted to a cutoff of
500~eV. The Brillouin-zone (BZ) integrations are carried out by means of
$\Gamma$-centered 12$\times$12$\times$$M$ {\bf k}-point meshes according to Monkhorst and Pack
\cite{Monkhorst.Pack:1976:PRB}. The value of $M$ is varied according to the number of bilayers $p=2,3,4,6$
(see Fig.~\ref{fig1}) in the unit cell. Explicitly, we use $M=6, 4, 3, 2$ for the 2$H$, 3$C$, 4$H$, and 6$H$ phase. We apply the GGA PBEsol XC functional \cite{Perdew.Ruzsinszky.ea:2008:PRL}, a modified version of the
Perdew-Becke-Ernzerhof (PBE) functional \cite{Perdew.Burke.ea:1996} optimized for solids, for structural optimizations. 
The lattice relaxation leads to a minimization of the total energy $E_{\rm tot}$ with a convergence accuracy of 1~meV/atom. The atomic geometry of the hexagonal polytypes $pH$ are characterized by lattice constants $a$ and $c$ and ($p-1$) internal-cell
parameters. The constants $a$ and $c$ give the cell volume $V=\sqrt{3/2}a^2c$ as well as the cell shape by their ratio
$c/a$. In the case of 3$C$ in Fig.~\ref{fig1} this ratio is fixed $2c/3a=\sqrt{8/3}$. The optimization of the positions of the atoms in the
unit cell is important as demonstrated for III-V compounds \cite{Panse.Kriegner.ea:2011:PRB}. Small deformations of the equilibrium atomic geometry are applied to extract elastic constants from the corresponding
total energy variations. The Murnaghan equation
of state (EOS) \cite{Murnaghan:1944:PNASU} $E_{\rm tot}=E_{\rm tot}(V)$ is applied to determine the equilibrium
volume $V_0$, the isothermal bulk modulus $B_0$, and its pressure derivative $B_0'$.

\subsection{\label{sec2b}Electronic states}

In order to compute approximate QP band structures we apply two approaches based on approximations of more advanced treatments of the XC
effects, the MBJLDA meta-GGA functional \cite{Becke.Johnson:2006:TJoCP,Tran.Blaha.ea:2007:JoPCM,Tran.Blaha:2009:PRL}
and the HSE06 hybrid functional \cite{Heyd.Scuseria.ea:2003,Heyd.Scuseria.ea:2006:JoCP}. Spin-orbit interaction is taken
into account for all calculations. To perform hybrid functional calculations we slightly reduce the density of the ${\bf k}$-point meshes to 8$\times$8$\times$$M$, where
$M=6$ for 2$H$, $M=4$ for 3$C$, $M=3$ for 4$H$ and $M=2$ for 6$H$ polytypes. In the case of calculations with the meta-GGA functional,  the cutoff energy was increased to 520~eV.
The eigenvalues of the parity operator were calculated using the code IrRep \cite{Iraola.Manes.ea:2022:CPC,Elcoro.Bradlyn.ea:2017:JAC}.

\section{\label{sec3}Results for ground-state properties}
\subsection{\label{sec3a}Structural parameters}

The calculated lattice parameters $a$ and $c$ of the hexagonal Bravais lattices (as well as of the diamond lattice in the hexagonal unit cell), together with their ratio
$c/a$ and the cell volume $V$ are listed in Table~\ref{tab1}.
Thereby the $c$ lattice constant is divided by the number of bilayers $p$ in the unit cell and the volume per  
atom is calculated as $V=\sqrt{3}a^2c/p$. We can observe clear trends with the hexagonality $h$, where $h=1.00$ for 2$H$, 0.50 for 4$H$, 0.33 for 6$H$ and 0.00 for 3$C$ polytypes, as indicated in Fig.~\ref{fig1}.

\begin{figure}[h]
\includegraphics[width=16cm]{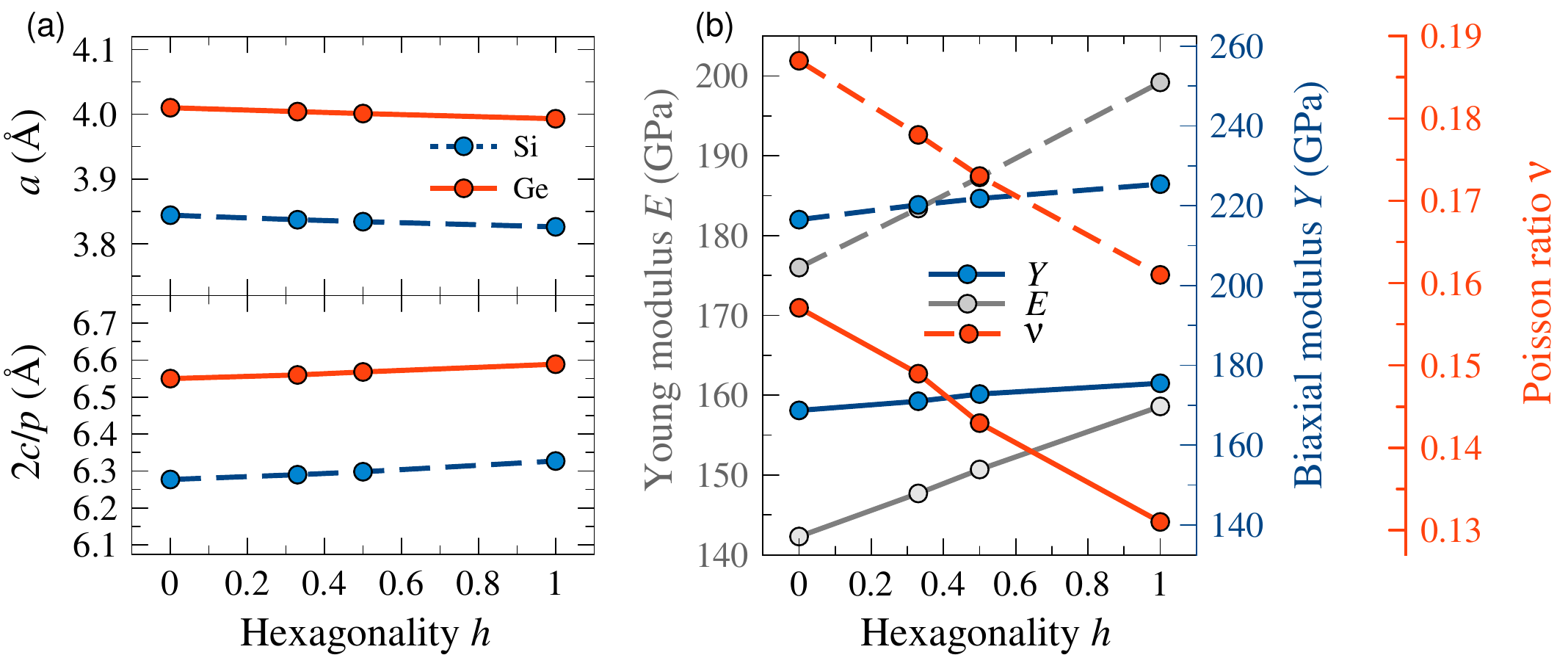} 
\caption{Dependence on hexagonality of (a) structural and (b) elastic properties of Si (dashed line) and Ge (solid line)
polytypes.}
\label{fig2}
\end{figure}

For both Si and Ge the lattice constant $a$ decreases, while the normalized lattice constant $2c/p$ and the ratio
$2c/(pa)$ increase with increasing hexagonality (see Fig.~\ref{fig2}(a)). The trend observed for $a$ also holds for the
volume per atom in the Ge case. However, for Si the trend is not monotonous, because of the opposite behavior of $a$
and $2c/p$. The cell volume $V$ reaches a minimum at $h=0.50$, i.e., for the 4$H$-Si polytype. The general trend with respect to $h$
indicates an increasing deformation of the tetrahedral units that are stretched along the $c$-axis.

\begin{table}[h!]
 \renewcommand{\arraystretch}{0.8}
\begin{scriptsize}
\caption{Equilibrium structural parameters and total energies (relative to the energy of the 3$C$ phase). A comparison with experimental
and other theoretical values is also given.
}
\centering
\scriptsize
\begin{ruledtabular}
\begin{tabular}{cccccccccc} 
Element & Polytype & $a$ ({\AA}) & $2c/p$ ({\AA}) & $2c/pa$ & $V$ ({\AA}$^3$/atom pair) & $B_0$ (GPa) & $B_0'$ & $E_{\rm tot}$ (meV/atom) & Reference \\ \hline
Si & 2$H$ & 3.826   & 6.327   & 1.654   & 40.109 & 93.5 & 4.25 & 9.7 & present \\
    &       &               &             &               &              &          &           &       & exp. \\
    &       & 3.824   & 6.257   & 1.6362 & 39.619 &         &          &          & \cite{Ahn.Kim.ea:2021:SSaT} \\
    &       & 3.8237 & 6.3237 & 1.6538 & 40.035 &         &         &           & \cite{Hauge.Verheijen.ea:2015:NL} \\
    &       & 3.837   & 6.317    & 1.646   & 40.271 &         &         &          & \cite{Besson.Mokhtari.ea:1987:PRL} \\
    &       & 3.840   & 6.280    & 1.630   & 40.098 &         &         &          & \cite{Zhang.Iqbal.ea:1999:APL} \\
    &       &               &             &               &              &          &         &          & theor. \\
    &       & 3.798   & 6.280    & 1.653   & 39.226  & 96.7 & 4.06 & 10.7 & \cite{Raffy.Furthmueller.ea:2002:PRB} \\
    &       & 3.828   & 6.325    & 1.652   & 40.133  & 92.8 & 4.24 &         & \cite{Roedl.Sander.ea:2015:PRB} \\
    &       & 3.800   & 6.270    & 1.650   & 39.204  &          &          & 11.7 & \cite{Yeh.Lu.ea:1992:PRB} \\
    & 4$H$ & 3.834   & 6.298    & 1.643   & 40.095  & 93.6 & 4.25 & 2.5  & present \\
    &       &               &             &               &              &          &           &       & exp. \\
    &       & 3.837   & 6.293    & 1.640   & 40.126 &         &         &          & \cite{Shiell.Zhu.ea:2021:PRL} \\
    &       & 3.840   & 6.270    & 1.633   & 40.034 &         &         &          & \cite{Pandolfi.ReneroLecuna.ea:2018:NL} \\
    &       &               &             &               &              &          &         &          & theor. \\
    &       & 3.806   & 6.254    & 1.643   & 39.228  & 96.7 & 4.13 & 2.4 & \cite{Raffy.Furthmueller.ea:2002:PRB} \\
    & 6$H$ & 3.837   & 6.290    & 1.639   & 40.106  & 93.5 & 4.25 & 1.2 & present \\
    &       &               &             &               &              &          &         &          & theor. \\
    &       & 3.810   & 6.244    & 1.639   & 39.248  & 96.7 & 4.13 & 1.0 & \cite{Raffy.Furthmueller.ea:2002:PRB} \\
    & 3$C$ & 3.844   & 6.277    & 1.633   & 40.152  & 93.5 & 4.25 & 0.0 & present \\
    &       &               &             &               &              &          &           &       & exp. \\
    &       & 3.840   & 6.272    & 1.633   & 40.047  & 97.9 & 4.24 & 0.0 & \cite{Madelung.Schulz:1987:Book} \\
    &       &               &             &               &              &          &         &          & theor. \\
    &       & 3.816   & 6.230    & 1.633   & 39.283  & 96.6 & 4.18 & 0.0 & \cite{Raffy.Furthmueller.ea:2002:PRB} \\ 
     \\ 
Ge & 2$H$ & 3.993  & 6.589   & 1.650   & 45.493 & 63.7 & 4.99 & 16.0 & present \\
    &       &               &             &               &              &          &           &       & exp. \\
    &       & 3.96      & 6.57    & 1.659    &              &         &          &          & \cite{Xiao.Pirouz:1992:JoMR} \\
    &       & 3.988    & 6.578  & 1.649    &              &         &          &          & \cite{Ziss.Stangl::Unpub} \\
    &       &               &             &               &              &          &         &          & theor. \\
    &       & 3.962   & 6.538    & 1.650   & 44.440  & 72.8 & 4.74 & 16.1 & \cite{Raffy.Furthmueller.ea:2002:PRB} \\
    &       & 3.996   & 6.590    & 1.649   & 45.566  & 67.6 & 4.81 & 10.0 & \cite{Roedl.Furthmueller.ea:2019:PRM} \\
    &       & 3.989   & 6.582    & 1.650   & 45.351  & 72.0 &          &         & \cite{Bao.Guo.ea:2021:JAP} \\
    & 4$H$ & 4.001    & 6.568   & 1.642   & 45.516  & 67.2 & 4.75 & 7.1  & present \\
    &       &               &             &               &              &          &           &       & exp. \\
    &       & 3.990    & 6.558  & 1.643    & 45.205 &         &          &          &  \\
    &       &               &             &               &              &          &         &          & theor. \\
    &       & 3.969   & 6.516    & 1.642   & 44.447  & 72.8 & 4.77 & 6.9 & \cite{Raffy.Furthmueller.ea:2002:PRB} \\
    & 6$H$ & 4.004    & 6.560   & 1.638   & 45.535  & 67.2 & 4.67 & 4.5  & present \\
    &       &               &             &               &              &          &         &          & theor. \\
    &       & 3.972   & 6.510    & 1.640   & 44.473  & 72.8 & 4.77 & 4.3 & \cite{Raffy.Furthmueller.ea:2002:PRB} \\

    & 3$C$ & 4.010   & 6.550    & 1.634   & 45.596  & 66.0 & 4.08 & 0.0 & present \\
    &       &               &             &               &              &          &           &       & exp. \\
    &       & 4.001   & 6.534    & 1.633   & 45.292  & 77.0 & 4.60 & 0.0 & \cite{Madelung.Schulz:1987:Book} \\
    &       &               &             &               &              &          &         &          & theor. \\
    &       & 3.979   & 6.496    & 1.633   & 44.534  & 72.5 & 4.80 & 0.0 & \cite{Raffy.Furthmueller.ea:2002:PRB} \\ 
\end{tabular}
\end{ruledtabular}
\label{tab1}
\end{scriptsize}
\end{table}

Our findings in Table~\ref{tab1} are in qualitative agreement with other calculations and experimental data. However,
systematic experimental studies are missing. The theoretical data obtained in the framework of DFT-LDA
\cite{Raffy.Furthmueller.ea:2002:PRB} show exactly the same qualitative behavior for Si and Ge, but also for SiC \cite{Kackell.Wenzien.ea:1994:PRBa}. The underestimation of $a$, $2c/p$ and $V$
in \cite{Raffy.Furthmueller.ea:2002:PRB} is a consequence of the well-known tendency to overbinding of the LDA functional 
\cite{Farid.Needs:1992:PRB}. The computed DFT-PBEsol lattice constants are much closer to the measured
$a$ and $c$ values, clearly indicating the improved quality of the PBEsol XC functional. For all hexagonal
polytypes the $2c/(pa)$ ratio is larger than the ideal value $\sqrt{8/3}=1.633$ of the 3$C$ polytype. This
result agrees with observations for III-V and II-VI compounds that crystallize in the zincblende structure
under ambient conditions \cite{Yeh.Lu.ea:1992:PRB,Bechstedt.Belabbes:2013:JoPCM}. By contrast,  the ratio $2c/(pa)$ of
compounds such as III-nitrides, SiC and some II-VI materials, for which the wurtzite 2$H$ polytype is more
stable than 3$C$, is below the ideal value
\cite{Bechstedt.Kackell.ea:1997:pssb,Yeh.Lu.ea:1992:PRB,Bechstedt.Belabbes:2013:JoPCM}.

Our results for the fits to the Murnaghan EOS are also displayed in Table~\ref{tab1}. The overall elastic properties
expressed by the isothermal bulk modulus $B_0$ and its pressure derivative $B_0'$ remain practically
uninfluenced by the polytype geometry in the case of Si. Because of the shallow Ge3$d$ core electrons, minor
deviations appear for Ge. The $B_0'$ values slightly increase with $h$, while $B_0$ exhibits a weakly
pronounced maximum for intermediate hexagonality values. Similar trends are observed using the
DFT-LDA approximation \cite{Raffy.Furthmueller.ea:2002:PRB}. However, due to the overbinding tendency of LDA,
the $B_0$ values are systematically larger by about $3-4\%$ in comparison to the DFT-PBEsol results.

\subsection{\label{sec3b}Energetics}

In Table~\ref{tab1} we list the total energy differences of the polytypes to the energy of the 3$C$ structure. We can see that  diamond 3$C$ is the most stable polytype
for both Si and Ge, in agreement with experiments. This crystal structure is followed by 6$H$ and 4$H$, whereas the lonsdaleite
geometry 2$H$ is substantially higher in energy. We conclude that the general trend is an increase of the internal energy 
with hexagonality. There is however a sustantial difference between Si and Ge: for the former the energy increase with hexagonality can be fitted with an exponential curve, while the growth is only approximately linear for Ge. The absolute values of the energy differences calculated with the PBEsol density functional are close to the DFT-LDA energies of Ref.~\cite{Raffy.Furthmueller.ea:2002:PRB}. Our result for the 2$H$-3$C$ energy difference for Si case is
also close to the value reported by Yeh et al.~\cite{Yeh.Lu.ea:1992:PRB}.

\begin{figure}[h]
\includegraphics[width=12cm]{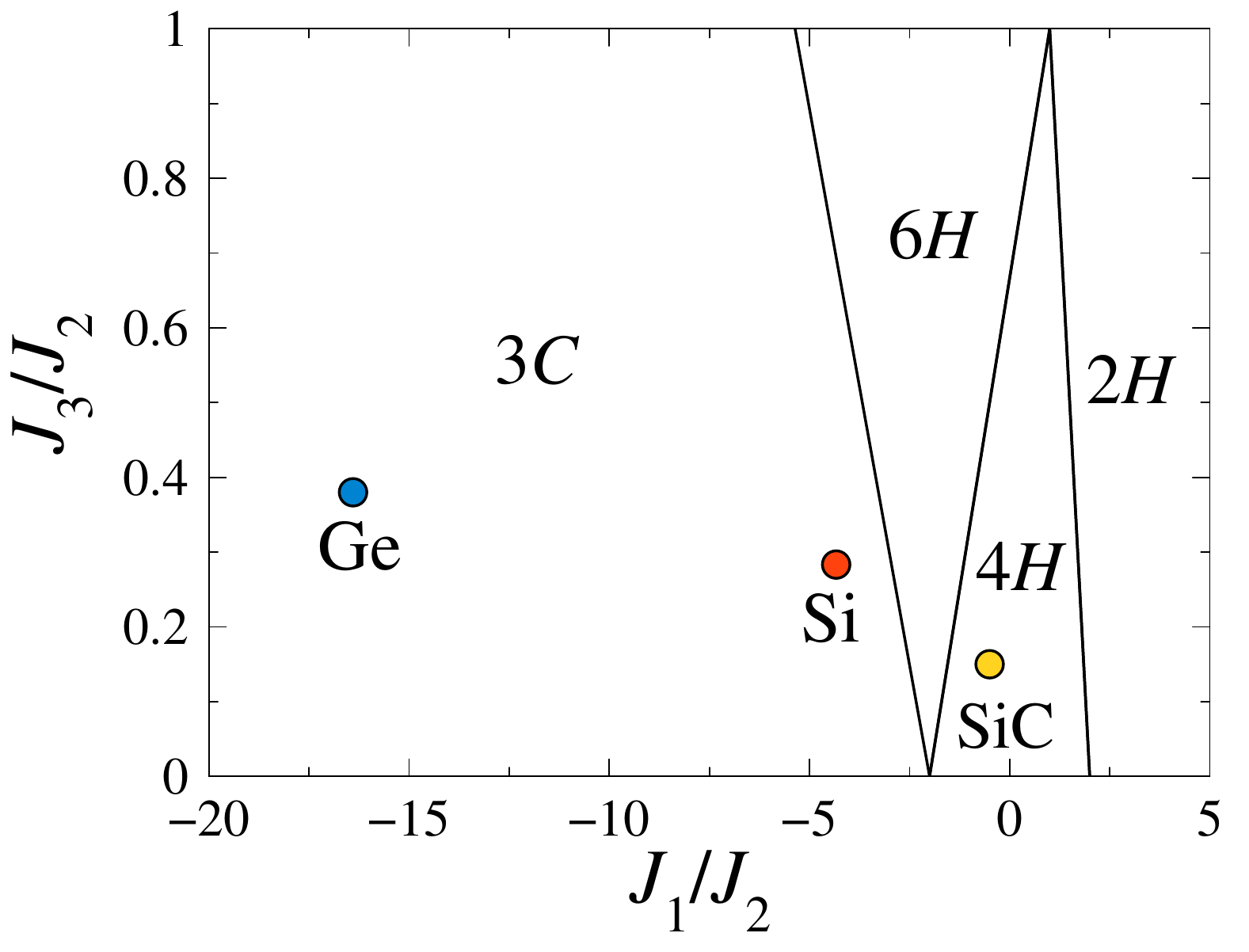} 
\caption{Phase diagram of polytypes from the ANNNI model. The solid lines represent the phase boundaries.
The two dots represent the group-IV materials. For comparison the 4$H$ equilibrium structure of
SiC \cite{Bechstedt.Kackell.ea:1997:pssb} is also displayed.}
\label{fig3}
\end{figure}

The polytypes differ only in the stacking sequence of the bonding tetrahedra along the $c$-axis. We can therefore model the internal energy of a generic polytype using 
a one-dimensional Ising-type model, called axial next-nearest-neighbor Ising (ANNNI) model \cite{vonBoehm.Bak:1979:PRL}, that uses three parameters $J_j$ $(j=1,2,3)$ to account for the $j$th-neighbor bilayer interaction. This model can be successfully 
applied to explain the energy differences in Table~\ref{tab1}
\cite{Bechstedt.Kackell.ea:1997:pssb,Raffy.Furthmueller.ea:2002:PRB,Bechstedt.Belabbes:2013:JoPCM,vonBoehm.Bak:1979:PRL,Cheng.Needs.ea:1988:JoPCSSP}.
The parameters of the model and the resulting stacking fault energies are summarized in Table~\ref{tab2}. The resulting phase diagram for 3$C$, 6$H$, 4$H$ and 2$H$ polytypes is shown in Fig.~\ref{fig3} together with the ratios of the ANNNI coefficients $J_1/J_2$ and $J_3/J_2$ for Si and
Ge.  This figure clearly shows that under ambient conditions Si and Ge crystallize in the diamond structure. However, the coordinate point $(J_1/J_2, J_3/J_2)$ is much closer for Si diamond than for Ge diamond to the 3$C$-6$H$ phase boundary with $J_1+2J_2+3J_3=0$ and
the triple point $J_1=-2J_2$ of the three polytypes 3$C$, 6$H$ and 4$H$. The reason is the larger (smaller)
nearest $J_1$ (second-nearest $J_2$) neighbor interaction in Ge (Si) (see Table~\ref{tab2}). The phase diagram
suggests that under near equilibrium conditions the preparation of hexagonal polytypes should be easier for Si. In fact the position of the diamond structure in the Si phase diagram is similar to the one of the ground state 4$H$ in the phase diagram of SiC, a compound which shows a pronounced polytypism \cite{Bechstedt.Kackell.ea:1997:pssb}.

Using the ANNNI model not only the different polytypes can be easily characterized. Also the formation
of a stacking fault, i.e., a 2D defect with respect to the infinite stacking in the 3$C$ structure can be studied with this simple model.
The most common stacking faults are the intrinsic stacking fault (ISF), the extrinsic stacking fault (ESF), and the twin stacking fault (TSF)
\cite{Mattheiss.Patel:1981:PRB,Chou.Cohen.ea:1985:PRB,Denteneer:1988:MOPL,Gross.Teichler:1991:PMB,Kaeckell.Furthmueller.ea:1998:PRB}.
The ISF (ESF) is defined by removing (adding) one bilayer from (to) the 3$C$ stacking sequence. A TSF defect
occurs if a reflection symmetry is present with the mirror plane positioned in the middle of the bilayer. The corresponding formation
energies (per atom) are given as \cite{Denteneer:1988:MOPL,Kaeckell.Furthmueller.ea:1998:PRB}
\begin{eqnarray}\label{eq1}
E_f(ISF)&=&4(J_1+J_2+J_3), \nonumber \\
E_f(ESF)&=&4(J_1+2J_2+2J_3), \\
E_f(TSF)&=&2(J_1+2J_2+3J_3). \nonumber
\end{eqnarray}
The stacking fault energies $\gamma$ per unit area can be calculated by dividing the formation energies of \eqref{eq1} by the area
$\sqrt{3}a^2/2$ of the hexagonal unit cell in the (111)/(0001) plane. The resulting values are
also listed in Table~\ref{tab2}.

\begin{table}[h!]
 \renewcommand{\arraystretch}{0.7}
\caption{Parameters $J_j$ of the ANNNI model (in meV/group-IV pair) and resulting stacking-fault energies
$\gamma$ (in mJ/m$^2$). Note that it is hard to extract precise data on the different types of stacking faults from experiments.
}
\centering
\footnotesize
\begin{ruledtabular}
\begin{tabular}{cccccccc} 
Element & $J_1$ & $J_2$ & $J_3$ & $\gamma_{\rm ISF}$ & $\gamma_{\rm ESF}$ & $\gamma_{\rm TSF}$ & Reference \\ \hline
Si & 10.4 & -2.4 & -0.68 & 36.5 & 21.2 & 8.9 & present \\
    &           &       &           &           &           &       & theor. \\
    & 11.4 & -2.9 & -0.75 & 39.3 & 20.6 & 8.4 & \cite{Raffy.Furthmueller.ea:2002:PRB} \\
    &           &       &           &           &           &       & exp. \\
    &            &       &           & 69   & 60  &    & \cite{Foell.Carter:1979:PMA} \\
    &            &       &           & 55$\pm$7   &    &    & \cite{Takeuchi.Suzuki:1999:pssa} \\ 
Ge & 16.4 & -1.0 & -0.38 & 69.4 & 63.3 & 30.8 & present \\
      &           &       &           &           &           &       & theor. \\
      & 16.5 & -1.2 & -0.46 & 69.7 & 62.2 & 30.0 & \cite{Raffy.Furthmueller.ea:2002:PRB} \\
      &           &       &           &           &           &       & exp. \\
    &            &       &           & 60$\pm$10   &    &    & \cite{Takeuchi.Suzuki:1999:pssa} \\ 
\end{tabular}
\label{tab2}
\end{ruledtabular}
\end{table}

These formation energies depend very weakly on the XC functional. However, they significantly depend on the chemical element. The theoretical values indicate that the stacking-fault generation is less energy-expensive in Si than in Ge, while the experimental formation
energies are rather similar. It is known from experiments that Si films crystallized with pulsed laser beams show many extrinsic stacking faults \cite{Kim.Lee:1997:TSF,Kim.Lee:1996:ML}.

\subsection{\label{sec3c}Internal-cell parameters}

The atomic positions in the unit cell of a $pH$ polytype are not only defined by the lattice constants $a$ and $c$, that fix the primitive basis vectors of the Bravais lattice $\bm{a}_1=a(1,0,0)$, $\bm{a}_2=\frac{a}{2}(-1,\sqrt{3},0)$
and $\bm{a}_3=c(0,0,1)$. The stackings in Fig.~\ref{fig1} can be described as ABABABABABAB$\ldots$ for 2$H$,
ABCABCABCABC$\ldots$ for 3$C$, ABCBABCBABCB$\ldots$ for 4$H$, and ABCACBABCACB$\ldots$ for 6$H$. The letters indicate the location of the vertical bonds (indicated with a black ticker line in Fig.~\ref{fig1} ) in the layers stacked along z in the unit cell.
The lowest atom sets the origin $(0,0,0)$, the other atoms in the vertical bonds have
Wyckhoff positions \cite{Wyckhoff:1964:Book} $(0,0,u)$,
$(1/3, 2/3,v)$, and $(2/3,1/3,w)$ with
\begin{eqnarray}\label{eq2}
2H &:& u_L=0, u_U=\frac{3}{8}+\varepsilon(1), \nonumber \\
4H &:& u_L=0, u_U=\frac{3}{16}+\varepsilon(1), v_L=\frac{1}{4}+\delta(2), v_U=\frac{7}{16}+\varepsilon(2), \\
6H &:& u_L=0, u_U=\frac{1}{8}+\varepsilon(1), v_L=\frac{1}{4}+\delta(2), v_U=\frac{7}{24}+\varepsilon(2),\nonumber \\
&& w_L=\frac{1}{3}+\delta(3), w_U=\frac{11}{24}+\varepsilon(3), \nonumber
\end{eqnarray}
where the deviations of the atomic positions from the ideal ones are given by the dimensionless internal-cell
parameters $\varepsilon(1),...,\varepsilon(p/2)$ for the upper ($U$) atom in the bilayer and $\delta(1)=0,...,\delta(p/2)$
for the lower ($L$) atom.

\begin{table}[h!]
 \renewcommand{\arraystretch}{0.7}
\caption{Internal-cell parameters for unit cells of type $pH$ in relative coordinates (fraction of the $c$ parameter, units of 10$^{-4}$) with respect to the lowest atom in the cell, whose internal parameter is set to zero. For comparison DFT-LDA values from Ref.~\cite{Raffy.Furthmueller.ea:2002:PRB}
are also listed.
}
\centering
\footnotesize
\begin{ruledtabular}
\begin{tabular}{cccccccccc} 
     & 2$H$ & \multicolumn{3}{c}{4$H$} & \multicolumn{5}{c}{6$H$} \\ \cline{2-2} \cline{3-5} \cline{6-10} 
Element & $\varepsilon(1)$ & $\varepsilon(1)$ & $\varepsilon(2)$ & $\delta(2)$ & $\varepsilon(1)$ & $\varepsilon(2)$ & $\varepsilon(3)$  & $\delta(2)$ & $\delta(3)$  \\  \hline
Si (this work)   &  \llap{$-$}10.0 & 6.6 &  \llap{$-$}1.7 & 8.3 & 6.2 & 3.3 &  \llap{$-$}1.2 & 7.4 & 2.9 \\
Ref.~\cite{Raffy.Furthmueller.ea:2002:PRB}      &  \llap{$-$}9.4   & 6.6 &  \llap{$-$}1.9 & 8.5 & 6.6 & 3.6 &  \llap{$-$}1.2 & 7.8 & 2.9 \\
Ge (this work) &  \llap{$-$}7.2   & 7.4 &  \llap{$-$}2.1 & 9.5 & 6.8 & 3.6 &  \llap{$-$}1.5 & 8.3 & 3.2 \\
 Ref.~\cite{Raffy.Furthmueller.ea:2002:PRB}     &  \llap{$-$}7.0   & 8.0 &  \llap{$-$}1.7 & 9.7 & 7.1 & 4.0 &  \llap{$-$}1.4 & 8.5 & 3.1 \\ 
\end{tabular}
\label{tab3}
\end{ruledtabular}
\end{table}

Table~\ref{tab3} shows that the geometry optimization produces only  small deviations from the ideal lattice positions. The discrepancies of the relative site positions between Si and Ge $pH$ polytypes are rather small, whereas the absolute
shifts are much larger because of the different values of the $c$-lattice constant (see Table~\ref{tab1}).
The values calculated with the PBEsol and LDA functionals are also very similar.

The largest deviations happen
for the 2$H$ londsdaleite polytype. However, even for 2$H$ the resulting $u=3/8+\varepsilon(1)$ parameter is still very close to the ideal value $u=0.375$, in agreement with previous computations (see e.g.
\cite{Roedl.Sander.ea:2015:PRB,Roedl.Furthmueller.ea:2019:PRM,Yeh.Lu.ea:1992:PRB}).

\subsection{\label{sec3d}Elastic coefficients}

The elastic stiffness constants $C_{ij}$ of the hexagonal polytypes $pH$ ($p=2,4,6$) and the 3$C$ polytype in a hexagonal unit cell are
extracted using DFT total energy calculations and the expression of the elastic energy for five different deformations
($<1\%$) of the crystal lattice. Combinations of $C_{ij}$ 
\cite{Kittel:2005:Book,Wang.Ye:2003:JoPCMa} yield the bulk modulus
\cite{Suckert.Roedl.ea:2021:PRM,Wagner.Bechstedt:2002:PRB}
\begin{equation}\label{eq3}
B_0=\frac{(C_{11}+C_{12})C_{33}-2(C_{13})^2}{(C_{11}+C_{12})+2(C_{33}-2C_{13})} \,,
\end{equation}
the Young modulus $E$, the biaxial modulus $Y$, and the Poisson ratio $\nu$
\cite{Suckert.Roedl.ea:2021:PRM,Wagner.Bechstedt:2002:PRB}
\begin{eqnarray}\label{eq4}
E&=&C_{33}-\frac{2(C_{13})^2}{C_{11}+C_{12}} \,, \nonumber \\
Y&=&C_{11}+C_{12}-\frac{2(C_{13})^2}{C_{33}} \,,\\
\nu&=&\frac{C_{33}}{C_{11}+C_{12}} \,.
\end{eqnarray}

\begin{table}[h!]
\caption{Elastic stiffness constants $C_{ij}$, bulk modulus $B_0$, Young modulus $E$, biaxial modulus $Y$
(all in GPa), and Poisson ratio $\nu$ (dimensionless). We compare the calculated values with reported theoretical values for 2$H$ and
experimental values for 3$C$.
}
\centering
\footnotesize
\begin{ruledtabular}
\begin{tabular}{ccccccccccccc} 
Element & Polytype & $C_{11}+C_{12}$ & $C_{11}$ & $C_{12}$ & $C_{13}$ & $C_{33}$ & $C_{44}$ & $B_0$ & $E$ & $Y$ & $\nu$ & Reference \\ \hline
Si & 2$H$ & 239.5 & 185.6 & 53.9 & 38.6 & 211.6 & 43.8 & 93.8 & 199.2 & 225.4 & 0.161 & present \\
    &       & 237.0 &            &          & 37.0 & 213.0 &           &         & 202.0 & 224.0 & 0.157 & \cite{Roedl.Sander.ea:2015:PRB} \\
    &       & 230.8 & 181.9 & 48.9 & 33.3 & 205.9 & 48.9 & 88.9 &            &            & 0.213 & \cite{Wei.Li.ea:2022:SMa} \\
    &       & 239.0 & 185.0 & 54.0 & 38.0 & 211.0 & 47.0 & 94.0 & 198.0 & 225.0 & 0.159 & \cite{Borlido:pc} \\
    &       & 249.3 & 194.0 & 55.3 & 42.0 & 206.5 & 44.8 & 97.0 & 192.3 & 232.2 & 0.168 & \cite{Wang.Ye:2003:JoPCMa} \\
    & 4$H$ & 238.8 & 183.4 & 55.4 & 41.4 & 201.7 & 50.9 & 93.9 & 187.3 & 221.8 & 0.173 & present \\
    &       &            & 179.9 & 50.2 & 36.0 & 197.1 & 52.2 & 89.1 &            &            & 0.214 & \cite{Wei.Li.ea:2022:SMa} \\
    & 6$H$ & 238.3 & 182.5 & 55.8 & 42.4 & 198.5 & 53.2 & 93.9 & 183.4 & 220.2 & 0.178 & present \\
    &       &            & 179.1 & 50.3 & 36.9 & 194.1 & 53.7 & 88.9 &            &            & 0.214 & \cite{Wei.Li.ea:2022:SMa} \\
    & 3$C$ & 237.0 & 183.4 & 53.6 & 44.4 & 192.7 & 56.0 & 93.8 & 176.0 & 216.5 & 0.187 & present \\
    &       & 248.8 & 191.4 & 57.4 & 44.8 & 204.0 & 57.9 & 97.9 & 203.6 & 248.4 & 0.180 & \cite{Wang.Ye:2003:JoPCMa} \\
Ge & 2$H$ & 182.5 & 143.1 & 39.3 & 23.9 & 164.9 & 40.1 & 69.5 & 158.6 & 175.5 & 0.131 & present \\
    &       & 177.7 & 124.0 & 53.7 & 22.8 & 159.4 & 39.1 & 67.3 & 153.5 & 171.2 & 0.128 & \cite{Suckert.Roedl.ea:2021:PRM} \\
    &       & 193.1 & 155.6 & 37.5 & 27.7 & 169.3 & 41.1 & 74.0 & 161.4 & 184.0 & 0.143 & \cite{Wang.Ye:2003:JoPCMa} \\
    &       & 179.0 & 138.0 & 41.0 & 25.0 & 161.0 & 38.0 & 77.0 & 154.0 & 171.0 & 0.140 & \cite{Borlido:pc} \\
    & 4$H$ & 181.3 & 141.2 & 40.0 & 25.9 & 158.1 & 42.0 & 69.4 & 150.7 & 172.8 & 0.143 & present \\
    & 6$H$ & 180.3 & 141.4 & 38.8 & 26.8 & 155.7 & 43.0 & 69.3 & 147.7 & 171.0 & 0.149 & present \\
    & 3$C$ & 179.3 & 142.7 & 36.6 & 28.2 & 151.2 & 44.9 & 69.2 & 142.3 & 168.7 & 0.157 & present \\
    &      & 183.3 & 154.2 & 35.3 & 23.3 & 159.9 & 47.6 & 68.9 & 154.0 & 176.5 & 0.127 & \cite{Martienssen.Warlimont:2005:Book} \\ 
\end{tabular}
\label{tab4}
\end{ruledtabular}
\end{table}

The calculated results are listed in Table~\ref{tab4}. We can compare the elastic constants of the lonsdaleite polytype with data computed recently within similar approaches and XC functionals \cite{Roedl.Sander.ea:2015:PRB,Borlido:pc,Wang.Ye:2003:JoPCMa}.
In the case of 2$H$-Ge, calculations \cite{Suckert.Roedl.ea:2021:PRM} performed with the ELASTIC
code \cite{Liu:2020:arxiv}, based on VASP total energies, are also available. For 3$C$ we re-express the stiffness constants \cite{Martienssen.Warlimont:2005:Book}
measured for the cubic system to obtain the coefficients corresponding to the hexagonal symmetry applying the formulas (including corrections)
given in Ref. \cite{Fuller.Weston:1974:JoAP}. Apart from the almost constant isothermal bulk modulus $B_0$ and
the hardly varying stiffness constant $C_{12}$, the other elastic constants show clear trends with hexagonality $h$. We observe an increasing trend with hexagonality $h$ for $C_{11}+C_{12}$, $C_{11}$, $C_{33}$, $E$ and $Y$ and a decreasing trend for $C_{13}$,
$C_{44}$ and $\nu$. The almost linear dependence with respect to the parameter $h$ is displayed in Fig.~\ref{fig2}(b)
for $E$, $Y$, and $\nu$. The dependence of the crystal stiffness on hexagonality is related to the different tetrahedron deformation and  stacking. Of course, in
comparison, variations of elastic properties due to the presence of a different chemical element, Si or Ge, are much stronger than effects related to the different polytype. This is especially
visible in the values of the bulk modulus $B_0$ or of the reciprocal compressibility, which varies slightly around 94~GPa for Si and
 68~GPa for Ge. A similar effect of the chemistry is visible for the other elastic properties. Only the Poisson ratios $\nu=0.161-0.187$ (Si) and $\nu=0.131-0.157$ (Ge) versus decreasing $h$
exhibit similar chemical and crystallographic ranges of variation. 

For the polytype 2$H$ the results can be
compared with data from previous calculations
\cite{Roedl.Sander.ea:2015:PRB,Suckert.Roedl.ea:2021:PRM,Borlido:pc,Wang.Ye:2003:JoPCMa}. We could verify an excellent agreement
 with stiffness constants and elastic moduli computed using the same or
 similar XC functionals in the GGA. If LDA functionals \cite{Wang.Ye:2003:JoPCMa}
are employed, the elastic constants are systematically larger, in agreement with the tendency for overbinding of LDA. For a detailed comparison all values can be found in Table~\ref{tab4}.

Experimental data are available for the 3$C$ polytype \cite{Martienssen.Warlimont:2005:Book}. We find
qualitative agreement between measured and computed values, with overall smaller theoretical values, as expected due to the slight tendency to underbind of the PBEsol functional. Very recently, combining nanoindentation and \emph{in situ} high-pressure
synchrotron X-ray diffraction, the Young modulus $E=152.4$~GPa, the bulk modulus $B_0=91.8$~GPa and
a Possion ratio $\nu=0.22$ of hexagonal silicon have been determined  \cite{Liang.Xiong.ea:2022:SM}. The deviations of $E$ and $\nu$ to
the calculated values in Table~IV may be traced back to the polycrystalline nature of the samples.

\section{\label{sec4}Results for excited-state properties}
\subsection{\label{sec4a}Band structures}

\begin{figure}[h]
\includegraphics[width=14cm]{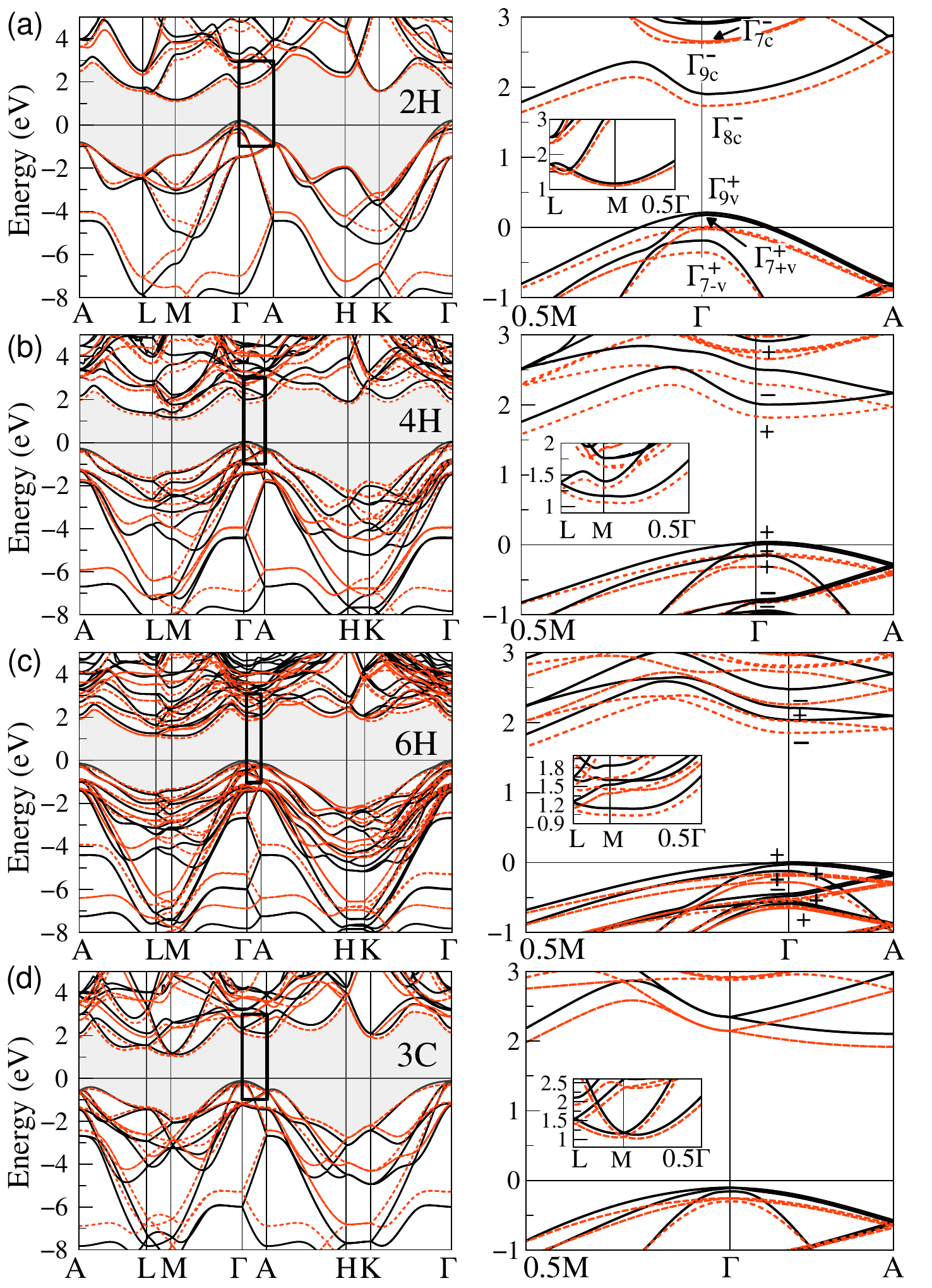} 
\caption{\scriptsize Approximate QP band structures of the four Si polytypes (a) 2$H$, (b) 4$H$, (c) 6$H$, and (d) 3$C$ along high-symmetry lines,
calculated within MBJLDA (red dashed lines) and HSE06 (black lines). The right panels show zooms on an energy interval around the fundamental gap and the BZ center, as indicated in the left panels by a black rectangle.  The BP is used as energy zero and indicated by a black horizontal line.
The insets show the lowest conduction bands in MBJLDA and HSE06 on the $\Gamma$-$M$-$L$ lines. The irreducible representations
 of high-symmetry states around the lowest conduction bands and highest valence bands are given in the double-group notation of
 Koster et al. \cite{Koster.Dimmock.ea:1963:Book} for 2$H$. Because of zone-folding arguments the denotation of 2$H$ is also applied in the cases of 4$H$ and 6$H$.
 \emph{Ab initio} band parities at $\Gamma$ are also displayed for all hexagonal polytypes. The gap regions are shaded in gray.}
\label{fig4}
\end{figure}

\begin{figure}[h]
\includegraphics[width=14cm]{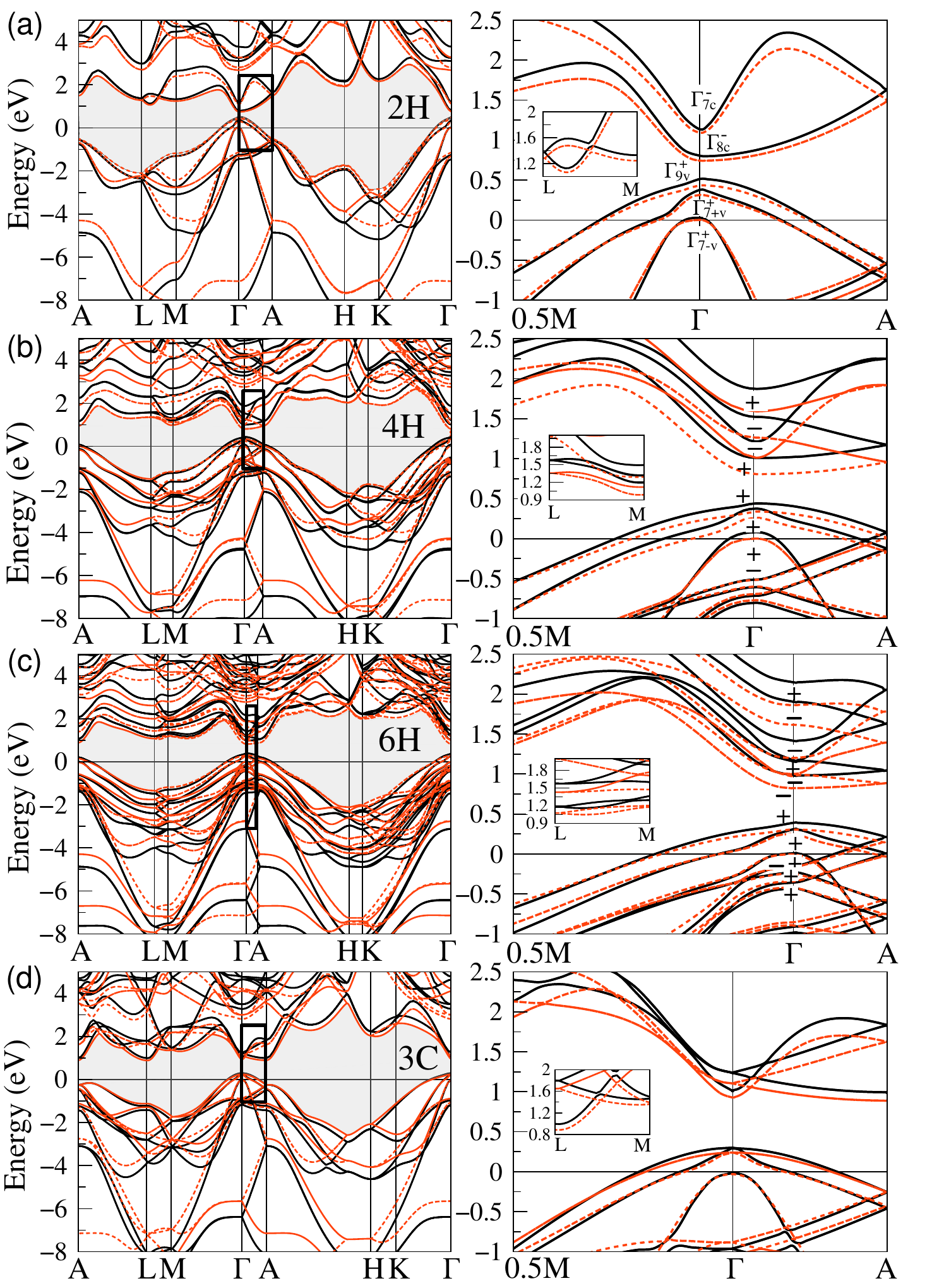} 
\caption{\scriptsize Approximate QP band structures of the four Ge polytypes (a) 2$H$, (b) 4$H$, (c) 6$H$, and (d) 3$C$ along high-symmetry lines,
calculated within MBJLDA (red dashed lines) and HSE06 (black lines). The right panels show zooms on an energy interval around the fundamental gap and the BZ center, as indicated in the left panels by a black rectangle.  The BP is used as energy zero and indicated by a black horizontal line.
The insets show the lowest conduction bands in MBJLDA and HSE06 on the $M$-$L$ line. The irreducible representations
 of the high-symmetry states around the lowest conduction bands and highest valence bands are given in the double-group notation of
 Koster et al. \cite{Koster.Dimmock.ea:1963:Book} for 2$H$. Because of zone-folding arguments the denotation of 2$H$ is also applied in the cases of 4$H$ and 6$H$.
 \emph{Ab initio} band parities at $\Gamma$ are also displayed for all hexagonal polytypes. The gap regions are shaded in gray.}
 \label{fig5}
\end{figure}

In Figs.~\ref{fig4} and \ref{fig5} we display the electronic band structures of the four considered polytypes of Si and Ge. The electronic states are calculated using approximate QP frameworks, namely the MBJLDA potential and the hybrid HSE06 functional. In the left panels the band energies are
plotted along the high-symmetry lines $A$-$L$-$M$-$\Gamma$-$A$-$H$-$K$-$\Gamma$ of the hexagonal
BZ. In the right panels one can observe zooms on the ${\bf k}$-space region around $\Gamma$, along the $\Gamma A$ and $\Gamma M$ directions, and on energies around the gap. The insets display the lowest conduction bands outside $\Gamma$.
For comparison, we also show the band structure of the cubic polytype, folded in the hexagonal BZ that results
from the use of a non-primitive hexagonal 3$C$ unit cell, as illustrated in Fig.~\ref{fig1}. The height of the different hexagonal
BZs varies with the polytype, i.e., it is determined by the number of bilayers $p$, along the $c$-axis, whereas the in-plane hexagonal sections perpendicular to the $c$-axis are basically equal. 

In order to align the band structures of the different polytypes we define the branch point (BP) energy as the common energy zero. We compute the BP for each polytype applying an approximate treatment
\cite{Schleife.Fuchs.ea:2009:APL} that has been reliably tested to give excellent results for band discontinuities
between semiconductors, their polytypes and their alloys
\cite{Schleife.Fuchs.ea:2009:APL,Bechstedt.Belabbes:2013:JoPCM,Belabbes.Bechstedt:2022:PRB}.
Following the procedure of Schleife et al.\cite{Schleife.Fuchs.ea:2009:APL} we use 2$p$ conduction and 4$p$ valence bands to construct the approximate charge neutrality point. If only half of the bands are applied the BP is shifted toward higher 
energies (see 2$H$-Si and 2$H$-Ge in \cite{Belabbes.Bechstedt:2022:PRB}). Resulting eigenvalues, band parameters, and gaps are
listed in Tables~\ref{tab5} and \ref{tab6}.
\begin{sidewaystable}
\caption{Energy eigenvalues (in eV) of the six lowest conduction ($c$) and three highest valence bands
($v$) at $\Gamma$ obtained from MBJLDA and HSE06 (values displayed in parenthesis) calculations. Each level is twofold degenerate. The state parity ($+/-$) is also listed. The BP is used as energy
zero. Its energy distance to the VBM can be obtained from the eigenvalues of the highest valence band $v_{1}$.
}
\centering
\scriptsize
\begin{ruledtabular}
\begin{tabular}{ccccccccc} 
          & \multicolumn{4}{c}{Si polytypes} & \multicolumn{4}{c}{Ge polytypes} \\ \cline{2-5}  \cline{6-9}   \\
Band &                   2$H$                 & 4$H$                             & 6$H$                            & 3$C$        &                       2$H$               &           4$H$                &             6$H$               & 3$C$ \\ \hline
$c_{6}$  & 4.835 ($-$)  (5.249)    &   3.015($-$)  (3.303)    & 2.847 ($-$)  (3.123)   &   3.881 (4.200)   &   4.946($-$)  (5.580)   &   2.910($-$)  (3.448) & 1.862 ($+$) (2.142)           & 3.003 (3.496)   \\
$c_{5}$ &  4.329 ($-$)  (4.780)    &   2.817 ($-$) (3.089)   & 2.823 ($-$) (3.098)    &   2.946 (3.231)    &   3.275($-$) (3.765)   &  2.758($-$)  (3.281) & 1.616 ($-$)  (1.899)            & 3.003 (3.491)   \\
$c_{4}$ &  3.014 ($-$)  (3.297)    &  2.794 ($-$) (3.063)  & 2.718 ($-$) (2.977)     &   2.946 (3.219)    &  2.811 ($-$)(3.251)   & 1.592 ($+$) (1.872) & 1.200 ($-$) (1.409)               & 2.807 (3.247)           \\
$c_{3}$ &  2.656 ($-$)  (2.930)    &  2.669 ($+$) (2.910)  & 2.256 ($-$) (2.476)   & 2.911 (3.186)   &  2.678 ($-$)  (3.104)    & 1.262 ($-$)  (1.522) & 1.001 ($+$) (1.168)               & 1.106 (1.238)   \\
$c_{2}$ &  2.634 ($-$) (2.902)    &  2.292 ($-$) (2.498)    & 2.030 ($+$) (2.210)  & 2.160 (2.346)    & 1.092 ($-$) (1.135)   & 1.011 ($-$)  (1.212) & 0.976 ($-$) (1.155)                  & 1.105 (1.238)   \\
$c_{1}$ & 1.735 ($-$) (1.903)    &  1.830 ($+$) (2.018)    & 1.860 ($-$) (2.035)    & 2.160 (2.345)    & 0.743 ($-$) (0.797)   & 0.807 ($+$)  (1.009) & 0.824 ($+$) (0.976)                & 0.930 (1.015)   \\
$v_{1}$ & 0.008 ($+$) (0.212)  &  -0.126 ($+$) (0.042)  & -0.160 ($+$) (0.004) & -0.255 (-0.102) & 0.435 ($+$) (0.514)  & 0.338 ($+$)  (0.441) & 0.306 ($+$) (0.390)                  & 0.240 (0.300)   \\
$v_{2}$ & -0.024 ($+$) (0.178) &  -0.154 ($+$) (0.010)  & -0.187 ($+$) (-0.024) & -0.255 (-0.102) & 0.315 ($+$) (0.379)  & 0.261 ($+$)  (0.376) & 0.251 ($+$) (0.308)                  & 0.236 (0.300)  \\
$v_{3}$ & -0.356 ($+$) (-0.146) &  -0.301 ($+$) (-0.154)  & -0.286 ($+$) (-0.122) & -0.302 (-0.156) & 0.002 ($+$) (0.028) & 0.001 ($+$) (0.083) & -0.004 ($+$) (0.008)               & -0.032 (-0.016) 

\label{tab5}
\end{tabular}
\end{ruledtabular}
\end{sidewaystable}

The energy levels can be labelled using the notation derived for lonsdaleite \cite{Koster.Dimmock.ea:1963:Book}. For the refinement of the representations we follow
Refs.~\cite{Joannopoulos.Cohen:1973:PRB,Joannopoulos.Cohen:1973:PRBa}, where a conduction
(valence) band state has the subscript $c$ $(v)$. All zone center states in 2$H$ lonsdaleite with space
group $D^4_{6h}$ belong to $\Gamma_7$, $\Gamma_8$ or $\Gamma_9$ representations with
either even or odd parity because of the center of inversion symmetry.
For all hexagonal polytypes 2$H$, 4$H$ and 6$H$ the band parities have been calculated as expectation values of the parity operator. They agree with
other calculated values for 2$H$-Ge \cite{Roedl.Furthmueller.ea:2019:PRM}, but disagree with the parities
derived within the empirical pseudopotential method (EPM) for both 2$H$-Si and 2$H$-Ge
\cite{De.Pryor:2014:JoPCM}. The irreducible representations of the space group at $\Gamma$
are just the representations of the point group $D_{6h}$, i.e., all symmetry operations of the
point group $C_{6v}$ as well as the inversion. For ${\bf k}$-points out of $\Gamma$ with a
finite component along the $c$-axis the little group is $C_{6v}$, i.e., a point on the $\Delta$
line, but recovers the point group $D_{6h}$ at the $A$ point. The little groups of the high-symmetry
${\bf k}$-points on the zone boundaries of the hexagonal BZ are $D_{3h}$ for $K$ and $H$, but
$D_{2h}$ for $L$ and $M$. Even including SOC, the bands in Figs.~\ref{fig4} and \ref{fig5} remain
twofold degenerate because of the inversion symmetry. Point group operations must be
followed by appropriate fractional translations to obtain the irreducible representation of a
wave function. Interestingly, at $A$ and $L$ but also along the $LM$ line couple of bands merge, so that, considering spin, a fourfold degeneracy appears. In the cubic case 3$C$ we do not apply band labels of the 2$H$ space group, despite the $O^7_h$ space group of the diamond geometry.
Information on the irreducible representations of the band states of 2$H$ considering
the hexagonal crystal field and SOC can be found elsewhere \cite{De.Pryor:2014:JoPCM}.

The lonsdaleite 2$H$ band structures are displayed in Figs.~\ref{fig4}(a) and \ref{fig5}(a). The
double-group notations of the irreducible representations of the band states are chosen
according to Refs.~\cite{Roedl.Furthmueller.ea:2019:PRM,De.Pryor:2014:JoPCM} including
the parity. The corresponding single-group notations for lonsdaleite without SOC can be found
elsewhere \cite{Joannopoulos.Cohen:1973:PRBa,Salehpour.Satpathy:1990:PRB,Murayama.Nakayama:1994:PRB}.
Rules for the transition between single- to double-group notations, i.e., without and with
SOC, for the $C^4_{6v}$ symmetry are listed in Refs.~\cite{Tronc.Kitaev.ea:1999:pssb,Kitaev.Tronc:2001:PRB}.
The most important band energies at $\Gamma$ are made visible in the zoomed band structures of Figs.~\ref{fig4}(a) and \ref{fig5}(a). They are listed
together with the energies of the CBM on the $\Gamma ML$ lines for Si and the $LM$ line for Ge
in Table~\ref{tab5}. The 2$H$-Si crystal remains an indirect-gap semiconductor with the CBM near $M$ and gap energy
$E^{\rm ind}_g(\Gamma^+_{9v}\rightarrow M_{5c})=1.10$ (0.98)~eV according to MBJLDA (HSE06) calculations. In the following all values will be given as MBJLDA (HSE06). While the lowest conduction band is a $sp$-derived $\Gamma^-_{8}$ state in both 2$H$ polytypes, the
weaker SOC and stronger chemical bonding give rise in Si to $p_{xy}$-type second and third conduction bands, 
$\Gamma^-_{9c}$ and $\Gamma^-_{7c}$ respectively, which only slightly split (by about 20 meV). In the case of Ge, instead,
the second $\Gamma^-_{7c}$ conduction band is mainly $s$-derived. Such pure $s$ band occurs for Si at the much higher
energy of 2.6~eV above the lowest conduction band. In 2$H$-Si the direct gap at $\Gamma$,
 $E^{\rm dir}_g(\Gamma^+_{9v}\rightarrow\Gamma^-_{8c})=1.73$ (1.69)~eV
is much larger. The second conduction band $\Gamma^-_{9c}$ lies 2.63 (2.72)~eV higher in energy. The situation is completely different in the 2$H$-Ge polytype, which becomes a direct-gap
semiconductor with $E^{\rm dir}_g(\Gamma^+_{9v}\rightarrow\Gamma^-_{8c})=0.31$ (0.29)~eV.
The second conduction band $\Gamma^-_{7c}=0.63$ (0.61)~eV and the CBM
$U_{5c}=0.62$ (0.62)~eV on the $LM$ line are somewhat higher in energy. 

The uppermost
valence bands $\Gamma^+_{9v}$, $\Gamma^+_{7+v}$ and $\Gamma^+_{7-v}$ are rather
similar in $2H$ Si and Ge. Only the larger SOC in Ge gives rise to larger energy splittings. Another interesting
high-symmetry point is $A$, because its possible mapping onto the $\Gamma$ point or the
$A\Gamma$ line in polytypes 4$H$, 6$H$ and 3$C$ with larger unit cells and, therefore, less extended
BZs in the direction of the $c$-axis. The uppermost split valence bands $A_{8v}+A_{9v}$ and $A_{7v}+A_{9v}$
lie in Si and Ge below the interesting energy region of $\Gamma^+_{9v}$, $\Gamma^+_{7+v}$ and
$\Gamma^+_{7-v}$. The lowest conduction band $A_{8c}+A_{7c}$, however, approaches the
energy region of the second-lowest conduction band $\Gamma^-_{9c}$ (Si) or $\Gamma^-_{7c}$
(Ge). It will therefore influence the conduction bands in 4H and 6H according to folding arguments.

\begin{table}[h!]
\caption{Important conduction and valence band splittings, distances of the $p$ lowest conduction bands, as well as direct and indirect gaps
between conduction and valence bands (in eV). Values are obtained from calculations using MBJLDA (HSE06).
}
\centering
\begin{ruledtabular}
\scriptsize{
\begin{tabular}{ccccccccc} 
Material & Polytype & $\Delta_{cf}$ & $\Delta_{\rm SO}$ & $\Delta_{\rm SO\|}$ & $\Delta_{\rm SO\bot}$ & $\Delta\varepsilon_c$ & $E^{dir}_g$($\Gamma$) & $E^{ind}_g$ (near $M$)\\ \hline
 Si    & 2$H$ & 0.345 (0.330) & 0.050 (0.054) & 0.050 (0.054) & 0.053 (0.053) & 0.899 (1.029) & 1.728 (1.691) & 1.096 (0.984) \\
        & 4$H$ & 0.156 (0.170) & 0.048 (0.054) & 0.048 (0.050) & 0.047 (0.048) & 0.965 (1.045) & 1.955 (1.976) & 1.233 (1.131) \\
         & 6$H$ & 0.104 (0.104) & 0.048 (0.052) & 0.048 (0.050) & 0.048 (0.047) & 0.986 (1.088) & 2.021 (2.031) & 1.247 (1.145) \\
         & 3$C$ & 0.000 (0.000) & 0.047 (0.054)  & 0.047 (0.056) & 0.046 (0.048) & 0.752 (0.874) & 2.415 (2.447) & 1.291 (1.232) \\ 
 Ge      & 2$H$ & 0.274 (0.299) & 0.278 (0.317) & 0.278 (0.332) & 0.271 (0.320) & 0.349 (0.338) & 0.308 (0.283) & 0.629 (0.615)    \\
        & 4$H$ & 0.140 (0.145) & 0.274 (0.310) & 0.274 (0.280) & 0.271 (0.306) & 0.786 (0.863) & 0.469 (0.568) & 0.648 (0.750)     \\
       & 6$H$ & 0.093 (0.102) & 0.272 (0.314) & 0.272 (0.362) & 0.271 (0.317) & 1.038 (1.166) & 0.518 (0.586) & 0.743 (0.769)    \\
       & 3$C$ & 0.000 (0.000) & 0.275 (0.316) & 0.275 (0.323) & 0.267 (0.308) & 0.176 (0.223) & 0.690 (0.715) & 0.643 (0.690)
     
\end{tabular}
}
\label{tab6}
\end{ruledtabular}
\end{table}

The most striking feature of the 4$H$, 6$H$, and 3$C$ band structures in Figs.~\ref{fig4}(b), (c), (d)
and \ref{fig5}(b), (c), (d) is the increase of the number of bands according to the increase of
atoms in the unit cell. The bands surrounding the fundamental gap, e.g. the lowest
conduction and highest valence bands, qualitatively show a similar behavior as those of the 2$H$ polytypes.
The three uppermost valence bands around $\Gamma$ keep their symmetry, parity and dispersion,
independently of the polytype, since other valence bands cannot be folded in the same energy range. The lowest conduction band also maintains its similarity with the CBM at $\Gamma$
with $\Gamma^-_{8c}$ symmetry. The next conduction band minimum is found near $M$, along
the $ML$ or $M\Gamma$ line. In the case of Si, the indirect CBM moves from $M$ in 2$H$ phase toward
a position along the $\Gamma M$ line, with an increasing distance from $M$ with decreasing
hexagonality. In the case of Ge the indirect minimum remains on the $LM$ line and do not show a
unique trend with the hexagonality. This is mainly due to the mapping of the $L$ point of the 2$H$ crystal structure onto the
$M$ point in the 4$H$ structure, and that of the 2$H$ minimum near $\frac{1}{3}ML$ onto $M$ in the BZ of 6$H$.
The different behavior of the indirect CBM appears already in the diamond structure, where
it occurs on the $\Gamma X$ line near 0.8~$\Gamma X$ for Si,  but at the $L$ point for Ge,
where $X$ and $L$ are high-symmetry boundary points of the fcc BZ. The band folding when going from 2$H$ to 4$H$ structures, and further to 6$H$, is also visible in the energy range of the lowest conduction
bands near $\Gamma$. While in Ge the $\Gamma^-_{8c}$ and $\Gamma^-_{7c}$ band ordering 
of 2$H$ is also preserved for 4$H$ and 6$H$, in the two latter cases additional conduction bands appear
in the corresponding energy region. For the 4$H$ polytype, the two (SOC-split) lowest conduction bands at $A$ of 2$H$ appear in the energy region of the $\Gamma^-_{8c}$ and $\Gamma^-_{7c}$ bands. For 6$H$-Ge four
such bands mapped from the original $A\Gamma$ line can be found. Thereby, the lowest
band state at $\frac{2}{3}A\Gamma$ in 2$H$ is now folded onto $\Gamma$ at an energy around the
$\Gamma^-_{8c}$ and $\Gamma^-_{7c}$ bands. In any case, the lowest optical transition is parity-forbidden for the 4$H$ symmetry,
in contrast to 2$H$ and 6$H$. Therefore, 4$H$-Ge should be not suitable for active optoelectronic applications.

In general, MBJLDA and HSE06 bands agree well near $\Gamma$ and around the gap region. Farther away 
from the band gap the discrepancies become larger. For instance, while the three uppermost valence bands
show excellent agreement, the lower valence bands are located at lower energies in the case of HSE06. This is due to
the contribution of $s$ orbitals to the lower valence bands and the stronger localization of $s$ states using hybrid functionals
\cite{Roedl.Furthmueller.ea:2019:PRM}.

\subsection{\label{sec4b}Valence and conduction band parameters}

From the MBJLDA (HSE06) band structures in Figs.~\ref{fig4} and \ref{fig5} we can extract the band parameters
and gaps in Tables~\ref{tab5} and \ref{tab6}. The six lowest conduction and highest
three valence bands at $\Gamma$ for the $pH$ and 3$C$ polytypes are listed in Table~\ref{tab5}. As
energy zero we choose the BP. In the Si case the BP energy is slightly above the valence
band maximum (VBM) in MBJLDA. Its distance to the VBM slightly increases with decreasing hexagonality, ranging from
0.008~eV to 0.255~eV, in good agreement with other studies and references therein
\cite{Belabbes.Bechstedt:2022:PRB}. Using the HSE06 functional, the BP position varies around the VBM. For 100\% hexagonality it is
0.212~eV below the VBM, but 0.102~eV above in the 3$C$ case. For Ge the BP is below the VBM, independent 
of the polytype and the approximate QP description. Its distance decreases from 0.435 (0.514)~eV to 0.240 (0.300)~eV
 with decreasing hexagonality of the polytype. Also this result is in good agreement with predictions in the literature
for 2$H$- and 3$C$-Ge (see \cite{Schleife.Fuchs.ea:2009:APL} and references therein). The absolute band positions
at $\Gamma$ with respect to the VBM in Table~\ref{tab5} are in agreement with similar calculations for 2$H$-Ge \cite{Roedl.Furthmueller.ea:2019:PRM}.
However, also the agreement with EPM calculations for 2$H$-Si and 2$H$-Ge is excellent \cite{De.Pryor:2014:JoPCM},
despite the use of completely different calculation methods.

\begin{figure}[h]
\includegraphics[width=10cm]{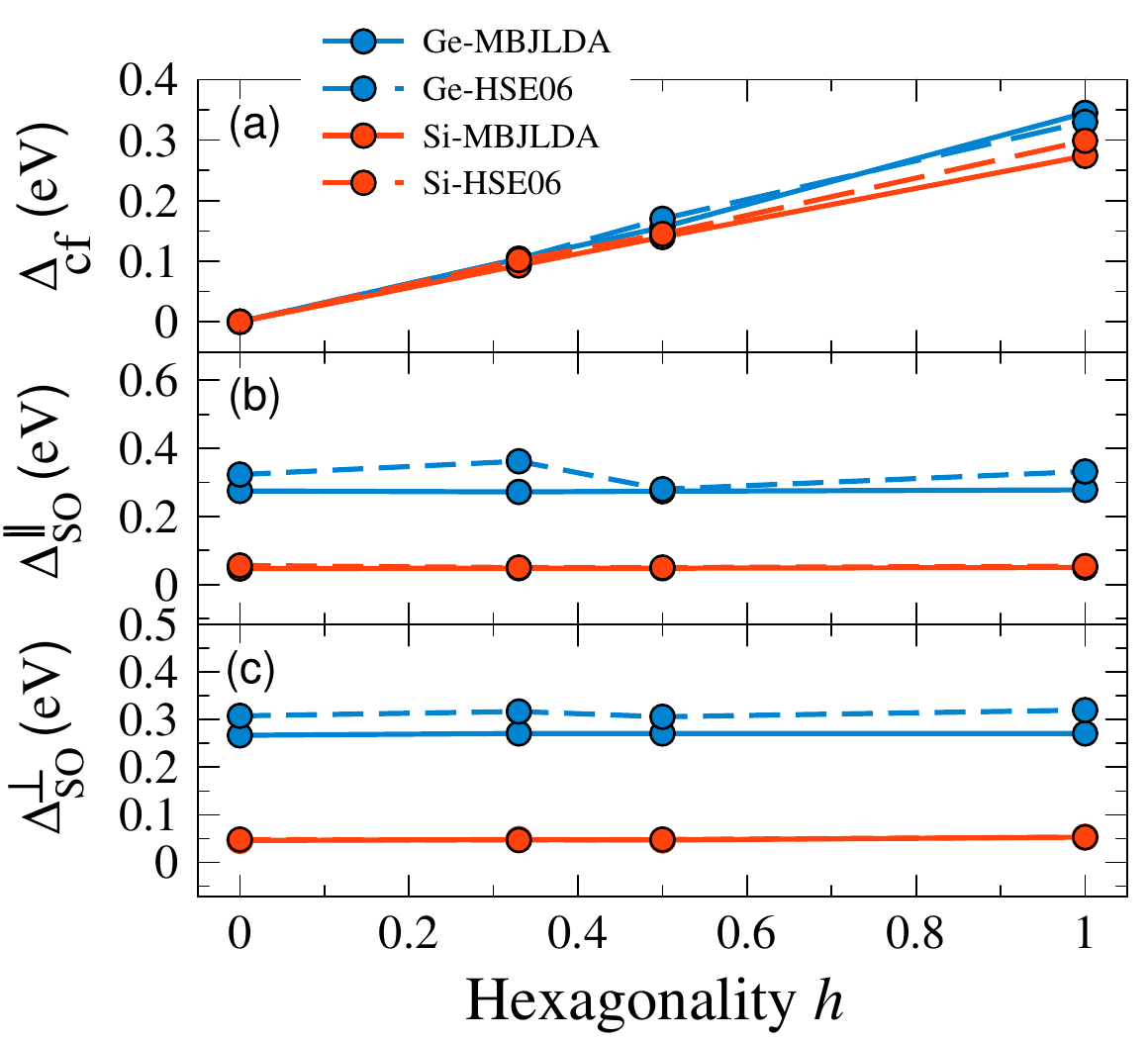} 
\caption{Valence band parameters (a) $\Delta_{cf}$, (b) $\Delta_{\rm SO\|}$ and (c) $\Delta_{\rm SO\bot}$ versus hexagonality of the
polytypes in MBJLDA (solid line) and HSE06 (dashed line) for Si (red) and Ge (blue), respectively.}
\label{fig6}
\end{figure}

From the valence band energies of $\Gamma^+_{9v}$, $\Gamma^+_{7+v}$, and $\Gamma^+_{7-v}$
in Table~\ref{tab5} we can extract the crystal-field splitting $\Delta_{cf}$ together with the spin-orbit splittings $\Delta_{\rm SO\|}$ and $\Delta_{\rm SO\bot}$ \cite{Schleife.Rodl.ea:2007}. To this purpose we apply formulas derived within the ${\bf k}\cdot{\bf p}$ theory
\cite{Chuang.Chang:1996:PRB}. The first values $\Delta_{cf}$ and $\Delta_{\rm SO}$ are derived within the quasicubic
approximation $\Delta_{\rm SO\|}=\Delta_{\rm SO\bot}=\Delta_{\rm SO}$ \cite{Bechstedt.Belabbes:2013:JoPCM,De.Pryor:2014:JoPCM}.
The more general formulas, accounting also for the hexagonal anisotropy, are presented elsewhere
\cite{Roedl.Furthmueller.ea:2019:PRM,Bechstedt.Belabbes:2013:JoPCM,De.Pryor:2014:JoPCM}.
The resulting values are listed in Table~\ref{tab6}. Their trends with the hexagonality $h$ of the
polytype is displayed in Fig.~\ref{fig6}. The hexagonal crystal field for the point group $D_{6h}$
leads to a splitting of the threefold (sixfold with spin) degenerate $\Gamma_{15}$ VBM of the
original diamond structure (without SOC) into the $p_{x,y}$-derived $\Gamma_6$ and the
$p_z$-derived $\Gamma_1$ levels, separated by $\Delta_{cf}$. The crystal field displays
an almost linear increase with the hexagonality, which is somewhat weaker for Si in comparison
with Ge. The spin-orbit interaction gives rise to a further splitting of $\Gamma_{6v}$ into
$\Gamma_{9v}$ and the $\Gamma_{7+v}$ states, while the mixing with $\Gamma_{1v}$ leads
to $\Gamma_{7-v}$. An additional splitting $\Delta_{\rm SO}$ occurs. In contrast to $\Delta_{cf}$ the SOC splitting
$\Delta_{\rm SO}$ hardly varies with the hexagonality, because of its strong atomic character. We observe only a tendency toward a weak
increase of the average SOC constant $\Delta_{\rm SO}$ with the polytype hexagonality. The absolute values of $\Delta_{cf}$
and $\Delta_{\rm SO}$ differ from those obtained with EPM for the 2$H$ polytypes
\cite{De.Pryor:2014:JoPCM}, because of the use of an ideal diamond-like atomic
symmetry in the latter case. The anisotropy of the hexagonal polytypes leads to the introduction
of two SOC constants $\Delta_{\rm SO\|}$ and $\Delta_{\rm SO\bot}$ \cite{Chuang.Chang:1996:PRB}.
In Table~\ref{tab6} we list the values resulting from the ${\bf k}\cdot{\bf p}$ formulas
\cite{Schleife.Rodl.ea:2007,Chuang.Chang:1996:PRB} by replacing $\Delta_{cf}$ with the value obtained in
the quasicubic approximation. These values are also plotted in Fig.~\ref{fig6}. In general, the hexagonal splitting $\Delta_{\rm SO\|}-\Delta_{\rm SO\bot}$ is
negligibly small and does not exhibit a unique trend with the hexagonality. In the case of 2$H$-Ge we find
an anisotropy splitting of 7~meV, in agreement with other computations \cite{Roedl.Furthmueller.ea:2019:PRM}. This 
splitting is of the order of 1 meV in the 2$H$ case.

Another interesting band parameter is the splitting of the $p$ lowest conduction bands
$\Delta\varepsilon_c$ at $\Gamma$, also listed in Table~\ref{tab6}. Along the row 2$H$,
4$H$ and 6$H$ there is a small increase with decreasing hexagonality for Si from 0.90 (1.03)~eV to
0.99 (1.09)~eV but a drastic variation from 0.35 (0.34)~eV to 1.04 (1.17)~eV for Ge. The 3$C$ band splitting 
 $\Delta\varepsilon_c=0.75$ (0.87)~eV (3$C$-Si) and 0.18 (0.22)~eV
(3$C$-Ge) do not follow the trends with hexagonality because of the more complex unfolding
behavior when we compare the lowest conduction bands in the fcc BZ and the non-primitive hexagonal BZ.

The direct gaps $E^{\rm dir}_g$ at $\Gamma$ in Table~\ref{tab6} also show a strong increase
with decreasing hexagonality from 1.73 (1.69) to 2.42 (2.45)~eV for Si and 0.31 (0.28) to 0.69 (0.72)~eV for Ge. This result is particularly
important for Ge because its hexagonal polytypes are pseudodirect semiconductors,
which have shown to be promising for applications in optoelectronics~\cite{Fadaly:2020:N}. Only the 3$C$ band structure displayed in the
hexagonal BZ, in agreement with its well known indirect character in the fcc BZ, exhibits a clear
indirect behavior with the CBM near $M$. The calculated indirect gap is $E^{\rm ind}_g=0.64$ (0.69)~eV.
These values are close to the 0.65 (0.68)~eV found for the indirect gap at the $L$ point of the fcc BZ
\cite{Roedl.Furthmueller.ea:2019:PRM}. In the Si case the indirect gap slightly increases with
decreasing hexagonality, going from 1.10 (0.98) to 1.29 (1.23)~eV in MBJLDA (HSE06). Interestingly, the true lowest direct gap of
3$C$-Si appears somewhat outside $\Gamma$ on the $\Gamma A$ line with a value of 2.17 (2.21)~eV.
Other HSE06 calculations \cite{Wei.Li.ea:2022:SM} for the Si polytypes yield very similar
values for the indirect gap values compared with those in Table~\ref{tab6}. Only the trend with the
hexagonality is not unique. The gaps of 2$H$-Si obtained from GW calculations
\cite{Roedl.Sander.ea:2015:PRB} are close to the values in Table~\ref{tab6}. Optical measurements
reveal that 4$H$-Si exhibits an indirect gap near 1.2~eV \cite{Shiell.Zhu.ea:2021:PRL}, in agreement with the
first-principles calculations. In the case of hexagonal Si nanoribbons an indirect gap of 1.5 eV has been measured~\cite{Qiu.Bender.ea:2015:SR5}. 

In the case of Ge other HSE06 or even GW calculations give a slightly smaller direct gap at
$\Gamma$, $E^{\rm dir}_g=0.23$~eV \cite{Fasolato.DeLuca.ea:2018:NL,Chen.Fan.ea:2017:JoPDAP}.
Band structure calculations with another hybrid exchange-correlation functional, B3LYP,
\cite{Kiefer.Hlukhyy.ea:2010:JMC} deliver a much larger direct gap of $E^{\rm dir}_g=0.81$~eV
for 4$H$-Ge. However, the B3LYP functional already tends to overestimate the gap of 3$C$-Ge.
Measurements of bulk, unstrained 2$H$-Ge are difficult. Photoluminescence measurements on
core-shell nanowires confirm a direct gap of about 0.3~eV \cite{Fadaly:2020:N}. Very recently
direct-band gap features have been also observed by photoluminescence for hexagonal Ge 
nanostructures \cite{Zhang.Yan.ea:2022:doi}. The measured direct gap of $E^{\rm dir}_g=0.8$~eV
has been related to atomically thin hexagonal layers embedded in cubic germanium. The resulting
carrier confinement and the compressive biaxial strain may explain the gap increase compared to 
the value of 0.3~eV calculated for a bulk, unstrained, hexagonal crystal.

\subsection{\label{sec4c}Band discontinuities}

Polytypic or heterocrystalline homojunctions have been observed growing Si or Ge nanowires
\cite{Fasolato.DeLuca.ea:2018:NL,Fabbri.Rotunno.ea:2014:SR,Vincent.Patriarche.ea:2014:NL}.
In particular, the 3$C$/2$H$ homojunction has been studied, also theoretically
\cite{Belabbes.Bechstedt:2022:PRB,Amato.Kaewmaraya.ea:2016:NL}. The alignment of the
band extrema by means of the BP energies in Table~\ref{tab5} allows to determine the natural band discontinuities $\Delta E_c$ and
$\Delta E_v$ for the homojunctions formed by two different polytypes of Si or Ge.

\begin{figure}[h]
\includegraphics[width=9cm]{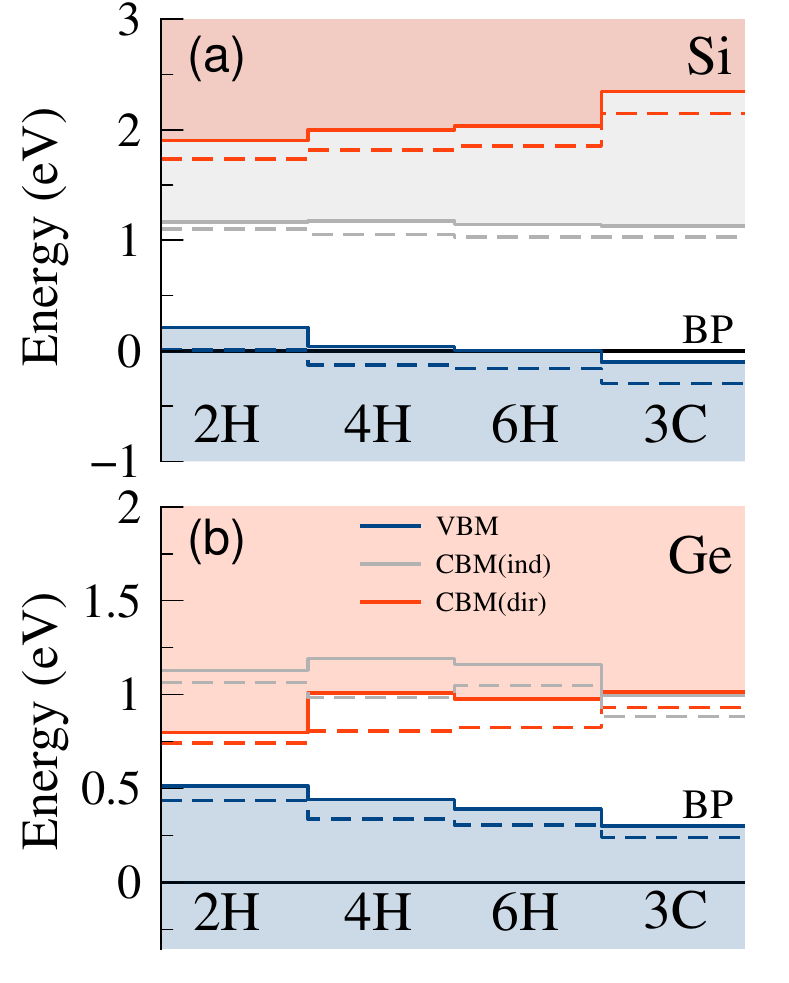} 
\caption{Conduction (direct in red and indirect in gray) and valence (blue) band edges of hexagonal Si (a) and Ge (b) polytypes aligned by their branch-point
energies (thin black horizontal line).  Dashed (solid) lines are computed using the MBJLDA (HSE06) functional.}
\label{fig7}
\end{figure}

In Fig.~\ref{fig7}(b) the situation for Ge is well defined because the enlargement of the direct
gaps along 2$H$, 4$H$, 6$H$ and 3$C$ is distributed over both band edges, CBM and VBM. A heterocrystalline
structure constituted by two different polytypes gives rise to a type-I heterostructure
\cite{Kittel:2005:Book}. The natural band discontinuities can be extracted from Table~\ref{tab5}:
$\Delta E_v=0.10$ (0.10), 0.03 (0.05), 0.06 (0.09), 0.19 (0.21)~eV and
$\Delta E_c=0.06$ (0.21), 0.02 (-0.03), 0.11 (0.04), 0.19 (0.22)~eV
for the junction 2$H$/4$H$, 4$H$/6$H$, 6$H$/3$C$, and 2$H$/3$C$, respectively.
The natural discontinuities $\Delta E_c=0.19$ (0.22)~eV and $\Delta E_v=0.19$ (0.21)~eV for the
2$H$/3$C$ junction are in close agreement with results in literature
\cite{Belabbes.Bechstedt:2022:PRB,Salehpour.Satpathy:1990:PRB,Fasolato.DeLuca.ea:2018:NL}, even if they vary slightly with
the numerical details of the calculation. The situation is less clear for the Si-based homojunctions
displayed in Fig.~\ref{fig7}(a). The natural band discontinuities
between the VBM are also well pronounced with $\Delta E_v=0.12$ (0.17), 0.03 (0.03), 0.09 (0.10), 0.26 (0.31)~eV for
2$H$/4$H$, 4$H$/6$H$, 6$H$/3$C$, and 2$H$/3$C$, while the conduction band minima are out of $\Gamma$
and, moreover, do only weakly vary. All the energies of the CBM are near the value 1.07 (1.15)~eV above the BP. Consequently, in contrast to the holes, the
electrons are hardly localized in real space in one of the considered Si-based homojunctions.
For the homojunction 2$H$/3$C$ we find $\Delta E_v=0.26$ (0.31)~eV and $\Delta E_c=-0.07$ ($-$0.04)~eV, i.e.,
a tendency for a type-II heterocharacter \cite{Kittel:2005:Book}. This qualitative finding is in
agreement with other theoretical predictions
\cite{Belabbes.Bechstedt:2022:PRB,Amato.Kaewmaraya.ea:2016:NL}.

\section{\label{sec5}Summary and conclusions}

We investigated the properties of hexagonal polytypes of Si and Ge, a new class of recently synthesized group-IV materials, using state-of-the-art \emph{ab initio} calculations. First, we discussed the structural, energetic, and elastic properties of the hexagonal polytypes 2$H$, 4$H$, and 6$H$ of Si and Ge, in comparison with those of the energetically most favorable 3$C$ diamond structure, obtained applying density functional theory with the exchange-correlation potential PBEsol . The crystal structures, including internal degrees of freedom, and the elastic properties show clear trends with hexagonality. We compare our results with calculations in literature performed with less accurate density functionals and with available measurements.

Some earlier surprising results on the energetics of the polytypes could be confirmed. The cubic 3$C$ polytype is certainly the most stable for Si and Ge. The total energy of the $p$H polytypes increases with increasing hexagonality. The increase is larger for Ge than for Si. Consequently, the phase diagram constructed within a generalized Ising model shows Si much closer to the triple point of 3$C$, 6$H$, and 4$H$ compared to Ge, indicating an easier production of the hexagonal polytypes in the case of Si, at least from the thermodynamic point of view. These results are consistent with the lower formation energies of stacking faults calculated for Si.

The different stacking of bonds in the polytypes affects the electronic properties. The direct (Si and Ge) gaps, as well as the indirect (Si only) gaps decrease with increasing hexagonality. There is a clear tendency for Si to be an indirect semiconductor, regardless of crystal structure, while hexagonal Ge polytypes are direct semiconductors. This rule is slightly broken in the case of 3$C$-Ge, where the lowest conduction band minima at $\Gamma$ and outside $\Gamma$ differs by only a few tenths of meV. The three uppermost valence bands are quite similar, regardless of the polytype.
Only the size of the crystal field ($\Delta_{cf}$) and spin-orbit splittings ($\Delta_{\rm SO}$) differ for Si and Ge. While $\Delta_{cf}$ varies almost linearly with the strength of the hexagonal crystal field, $\Delta_{\rm SO}$ remains fairly constant for all polytypes.

The band alignment between Ge polytypes is particularly interesting, in view of the direct band gap, for applications in optoelectronics. We find that all heterostructures constituted by two polytypes of Ge have a type-I character, with electrons and holes confined in the layer with higher hexagonality. These results can suggest design rules for quantum-well light emitters. 

\section*{Acknowledgements}

We acknowledge financial support from the EU through H2020-FETOpen project OptoSilicon
(Grant Agreement No. 964191).

\bibliography{sige}

 \newpage
 \section*{\label{figcap}Figure Captions}

\end{document}